\documentclass[aps,prb,twocolumn,superscriptaddress,showpacs,citeautoscript]{revtex4-1}

\usepackage{graphicx}  
\usepackage{dcolumn}   
\usepackage{bm}        
\usepackage{amssymb}   

\usepackage{pdfpages}
\usepackage{amsfonts}
\usepackage{amsmath}
\usepackage{color}
\usepackage{textcomp} 
\usepackage{dcolumn}
\usepackage{microtype}   
\usepackage{mathtools}

\usepackage[sort&compress]{natbib}
\newcommand{\bes}{\begin{split}}
\newcommand{\ees}{\end{split}}
\newcommand{\be}{\begin{equation}}
\newcommand{\ee}{\end{equation}}
\newcommand{\bea}{\begin{eqnarray}}
\newcommand{\eea}{\end{eqnarray}}
\newcommand{\ba}{\begin{array}}
\newcommand{\ea}{\end{array}}
\newcommand{\ball}{\begin{align}}
\newcommand{\eall}{\end{align}}
\newcommand{\biz}{\begin{itemize}}
\newcommand{\eiz}{\end{itemize}}

\newcommand{\Le}{\left}
\newcommand{\Ri}{\right}
\newcommand{\nn}{\nonumber}

\newcommand{\f}{\frac}

\newcommand{\mrm}[1]{\mathrm{#1}}

\newcommand{\bit}{\begin{itemize}}
\newcommand{\eit}{\end{itemize}}

\newcommand{\mc}{\mathcal}

\newcommand{\dg}{\dagger}
\newcommand{\ra}{\rangle}
\newcommand{\la}{\langle}
\newcommand{\Su}{|\uparrow\rangle}
\newcommand{\Sd}{|\downarrow\rangle}
\newcommand{\veps}{\varepsilon}

\newcommand{\ua}{\uparrow}
\newcommand{\da}{\downarrow}

\newcommand{\fkm}{f_{k}^{(m)}}
\newcommand{\fkn}{f_{k}^{(n)}}
\newcommand{\fn}[1]{f^{(#1)}}

\newcommand{\fa}[2]{f^{(#1)}_{#2}}

\newcommand{\spin}[1]{\sigma_{{#1}}}

\newcommand{\bkd}{b^{\dagger}_{k}}
\newcommand{\bk}{b^{\phantom{\dagger}}_{k}}

\newcommand{\akd}{a^{\dagger}_{k}}
\newcommand{\ak}{a^{\phantom{\dagger}}_{k}}
\newcommand{\w}{\omega}

\newcommand{\ket}[1]{\left| #1 \right\rangle}
\newcommand{\bra}[1]{\left\langle #1 \right|}

\newcommand{\wc}{\omega_{\mathrm{c}}}

\begin{document}

\title{A generalized multi-polaron expansion for the spin-boson model: \\ Environmental entanglement and the biased two-state system}

\author{Soumya Bera}
\affiliation{Institut N\'{e}el, CNRS and Universit\'e Grenoble Alpes, F-38042
Grenoble, France}
\author{Ahsan Nazir}
\affiliation{Photon Science Institute \& School of Physics and Astronomy, University of Manchester, Oxford Road, Manchester M13 9PL,
United Kingdom}
\affiliation{Controlled Quantum Dynamics Theory, Imperial College London, London SW7 2AZ,
United Kingdom}
\author{Alex W. Chin}
\affiliation{Theory of Condensed Matter Group, University of Cambridge,
J J Thomson Avenue, Cambridge, CB3 0HE, United Kingdom}
\author{Harold U. Baranger}
\affiliation{Department of Physics, Duke University, Durham, North Carolina
27708, USA}
\author{Serge Florens}
\affiliation{Institut N\'{e}el, CNRS and Universit\'e Grenoble Alpes, F-38042
Grenoble, France}

\date{\today}

\begin{abstract}
We develop a systematic variational coherent state expansion for the many-body
ground state of the spin-boson model, in which a quantum two-level system is
coupled to a continuum of harmonic oscillators. Energetic constraints at the
heart of this technique are rationalized in terms of polarons (displacements of
the bath states in agreement with classical expectations) and antipolarons
(counter-displacements due to quantum tunneling effects). We present a
comprehensive study of the ground state two-level system population and
coherence as a function of tunneling amplitude, dissipation strength, and bias
(akin to asymmetry of the double well potential defining the two-state system).
The entanglement among the different environmental modes is investigated by
looking at spectroscopic signatures of the bipartite entanglement entropy
between a given environmental mode and all the other modes.  We observe a
drastic change in behavior of this entropy for increasing dissipation,
indicative of the entangled nature of the environmental states.  In addition,
the entropy spreads over a large energy range at strong dissipation, a testimony
to the wide entanglement window characterizing the underlying Kondo state.
Finally, comparisons to accurate numerical renormalization group calculations
and to the exact Bethe Ansatz solution of the model demonstrate the rapid
convergence of our variationally-optimized multi-polaron expansion, suggesting
that it should also be a useful tool for dissipative models of greater
complexity, as relevant for numerous systems of interest in quantum physics and
chemistry.
\end{abstract}

\maketitle

\section{Introduction}

The study of open quantum systems constitutes an important and active frontier
of research, with several difficult challenges to
overcome~\cite{BreuerPetruccione,Weiss,Leggett,NitzanBook}.  First, one faces
the quantum complexity that comes from the combination of a quantum sub-system
with a macroscopic reservoir of environmental states. The former may range from
a two-state system (for instance, an atom or a superconducting qubit) to a very
large object (e.g.\ a molecule), while the latter can describe a variety of
complex excitations (phonons, spin baths, mobile solvent species,
electromagnetic fluctuations, etc.).  A second difficulty is the emergence of
strong correlation physics when the coupling to the bath becomes sufficiently
large, leading to substantially renormalized physical properties for the quantum
sub-system as compared to its behavior ``in vacuum''. This effect is typically
accompanied by the emergence of a complex entanglement structure within the
environment itself, involving modes that can span several decades in
energy.~\cite{Bera} This kind of complex behavior can, for instance, be captured
by sophisticated numerical techniques, such as the numerical renormalization
group (NRG)~\cite{NRG-RMP08,Bulla,Florens_DissipSpinDyn11}, or the density
matrix renormalization group and its variational matrix-product states
extension~\cite{Verstraete_MPS08,Wong,Guo}. 
Third, much of the interesting physics in this context, such as decoherence and
relaxation to a steady state, arises in an out-of-equilibrium dynamical
situation. This pushes numerical methods even further toward their limits~\cite{
Makri95,Anders,WangThoss08,AlvermannFehske09,PriorPlenio_EffSim10,NalbachThorwart10,Orth,Yao,Peropadre},
and many studies have hitherto focused on simpler analytical
approaches~\cite{Weiss,Leggett} (master equations, for instance) that may fail
at large dissipation and low temperatures.

The purpose of this paper is to develop an alternative to time-consuming
numerically-exact techniques, which benefits from the simplicity of a
variational treatment, yet has the potential to be scaled up to the exact
solution of the problem. The basis of the method, presented in a recent
publication,~\cite{Bera} is an expansion of the complete wavefunction of the
bath into a set of coherent states, which has the freedom to capture the
correlations that typically build up in open quantum systems in an economical
way. 
We shall focus on the problem of the many-body ground state in open
quantum systems within the challenging regime of strong coupling to the bath,
while the question of time-dependent dynamics will be considered elsewhere.  It
has been recognized since the work of Silbey et al. and Emery et
al.~\cite{EmeryLuther,
EmeryLutherPRB, Silbey1,Silbey2, Chin_SubohmSBM11} that the use of a {\it
single} multi-mode coherent state describes the main classical structure of the
environment, and amounts to polaron formation. We showed recently that quantum
mechanics imposes a finer structure on the states of the bath,~\cite{Bera}
whereby classically forbidden displacements gradually emerge for low energy
environmental modes.  This leads to a rich entanglement structure that can be
captured already by including a second multi-mode coherent state, dubbed the
antipolaron.  Adding an increasing number of antipolarons allows rapid
convergence of several ground state observables. The physical role of these
antipolarons is crucial in determining the magnitude of the coherence in the
system, which somewhat counter-intuitively is \emph{enhanced} with respect to
the purely classical (single polaron) description of the bath.

Here, we shall further develop and explore the multi-polaron expansion. First,
we shall examine the structure of the displacements at large dissipation,
showing that for the additional polarons they always snake between polaron and
antipolaron values, leading to a complex nodal structure in their energy
dependence. We shall also show that the proliferation of antipolarons at
increasing dissipation is associated to emerging structures in the entanglement
entropy obtained by tracing out the whole bath apart from a single mode.  A
second aspect of our study will be to generalize the dissipative two-state model
to a finite bias (i.e.~the parameter controlling the asymmetry in the
double-well potential underlying the two-level system).  While the structure of
the system-plus-bath wavefunction becomes more complicated in this case due to
the presence of symmetry breaking terms, we shall show that it is again
primarily controlled by energetics.  These considerations are not only important
for a good understanding of the physics at play, but also crucial in order to
efficiently determine the optimal displacements 
that define the best
set of variational states.  All of our results will be compared to accurate NRG
calculations (in the unbiased case), and to the exact Bethe Ansatz solution
(with bias)~\cite{Ponomarenko,LeHurReview}, demonstrating the fast 
convergence of our multi-polaron expansion. This is also confirmed by
establishing that 
the energy variance of our variational
state rapidly vanishes as a function of the number of coherent states used,
showing that our trial state quickly becomes an exact eigenstate of the model.
These precise checks put the coherent state expansion on a firm mathematical
ground and should make it a practical tool for more advanced applications, for
instance to sub-Ohmic environments, multiple baths and qubits, or generalized
multi-state systems.

The paper is organized into two major parts. We begin in
Section~\ref{sec:unbiased} by reviewing the single-polaron Silbey-Harris (SH)
theory, and show that a simple two-polaron ansatz cures the pathologies
associated to the SH state. In particular, we elucidate how environmental
correlations originating from antipolaronic effects preserve the spin coherence
$\la\spin{x}\ra$ in the ground state. We further generalize the variational
technique to account for a complete multi-polaron expansion of the ground state
of the unbiased spin-boson model, demonstrating excellent agreement with NRG
calculations.  New kinds of displaced states emerge, involving two (and possibly
more) nodes in their momentum dependence. Incorporating many polaron states into
the trial wavefunction allows us to account for the progressive build up of
entanglement within the bath at increasing dissipation, and we propose an
entropy measure to characterize precisely this property.  Finally, we check that
the energy variance of the multi-polaron ground state drops rapidly to zero on
increasing the number of polarons.

Subsequently, in Section~\ref{sec:biased}, we study the ground state of the
biased spin-boson model.  Here, we extend the multi-polaron ansatz to
incorporate the asymmetry between different oscillator displacements due to
finite bias. We show that the multi-polaron displacements for the biased model
possess 
very similar features 
to the unbiased case, such as
the formation of low-energy antipolarons and the stabilization of spin coherence
in comparison  to the classical (purely polaronic) response.  Rapid convergence
to the exact Bethe Ansatz solution, valid in the scaling limit of small
tunneling amplitude, is also verified. 
We conclude the
manuscript with a brief discussion of possible further extensions of the
multi-polaron technique to several more challenging open problems.

\section{Unbiased spin-boson model}\label{sec:unbiased}

\subsection{Hamiltonian}

The unbiased spin-boson~(SB) Hamiltonian~\cite{Leggett} (setting $\hbar=1$),
\be 
\mc{H} = \f{\Delta}{2} \sigma_x  - \f{\sigma_z}{2}\sum_{k > 0} g_k (b_k +
b_k^{\dg}) + \sum_{k>0} \w_k b_k^{\dag} b_k,
\label{eq:SB}
\ee 
describes a two-level system with tunneling energy $\Delta$ coupled to a
bath of harmonic oscillators with modes of energy $\w_k$. Here
$b^{\dg}_k$ ($b_k$) are the standard creation (annihilation) operators for
bosonic modes with momentum $k$, and the Pauli matrices are introduced to describe the
two states of the sub-system, in complete analogy to a spin $1/2$ ($\sigma_z=|\uparrow\rangle\langle\uparrow|-|\downarrow\rangle\langle\downarrow|$). The effect of the interaction $g_k$ between 
the spin and the modes is encapsulated by the bath spectral density:
\be 
\label{eq:sden}
	J(\w) = \sum_{k>0} \pi g_k^2 \delta(\w-\w_k)
= 2 \pi \alpha \w  
	\theta(\wc-\w),
\ee
where $\alpha$ is a dimensionless measure of the interaction strength between the two-level
system and the environment, and $\wc$ is a high frequency cutoff for the spectrum, 
which is assumed to be Ohmic (linear in frequency) in all
that follows. In the continuum limit, the $k$-sum appearing in the above equations is understood
as a continuous integral, a nomenclature that we shall use throughout the paper.

\subsection{Review of Silbey-Harris variational theory: Single-polaron ansatz}
The motivation for the polaronic variational treatment can be understood easily
by first considering the zero-tunneling limit, $\Delta = 0$. In this case, the model [Eq.~(\ref{eq:SB})] 
reduces to a system of harmonic oscillators with finite displacements
$f_k=g_k/(2\w_k)$ and energy $E(\Delta=0) = -\sum_{k>0}g_k^2/(4\w_k)$
in the ground state.
This ground state is two-fold degenerate, as the two-level system may freely point 
up or down.
Conversely, when the two-level system is fully decoupled from the bath ($g_k = 0$), 
the spin admits a single ground state $(\Su-\Sd)/\sqrt{2}$ with energy 
$E(\alpha=0) = -\Delta/2$, and is thus delocalised over the two minima of the underlying 
double-well potential.
For finite tunneling and dissipation, the system exhibits an inherent
competition between localization, induced by the interaction with the environmental bosons, and
delocalization intrinsic to the spin tunneling process. 

Several authors have proposed and studied an approximate multi-mode coherent state 
ansatz~\cite{EmeryLuther,EmeryLutherPRB,Silbey1, Silbey2} for the unbiased spin-boson model, that can capture  
both limiting cases. We shall refer to this as the SH (Silbey-Harris) ansatz; it takes the form  
\be
|\Psi_\mrm{SH}\ra= \f{1}{\sqrt{2}}\Big[ \Su \otimes |+f^\mrm{SH}\ra -
\Sd \otimes |-f^\mrm{SH}\ra \Big]
\label{eq:SHansatz}, 
\ee
where $|f^\mrm{SH}\ra  =  e^{\sum_{k>0}f_k^\mrm{SH} (b_k^{\dg}-b_k)}|0\ra$
is a multi-mode coherent state. 
Here, $|0\ra $ represents the full vacuum state with all oscillators in their 
undisplaced configuration. The oscillator displacements $f_k^\mrm{SH}$ 
are determined from the variational principle, based on minimizing the 
energy resulting from the ansatz when applied to the spin-boson Hamiltonian [Eq.~\eqref{eq:SB}]:
\be
E_\mrm{SH} = - \f{\Delta}{2} e^{-2 \sum_{k>0}(f_{k}^\mrm{SH})^2} + \sum_{k>0}
\w_k (f_{k}^\mrm{SH})^2 - \sum_{k>0} g_k f_{k}^\mrm{SH}. 
\label{eq:EnSh}
\ee
The first term in Eq.~\eqref{eq:EnSh} is the renormalized spin tunneling energy,
and the last two terms represent a shifted parabolic potential for the
oscillators.
The variational energy is minimized  according to 
$\partial E_{\rm SH}/\partial f_k^\mrm{SH} = 0$, leading to optimal displacements,  
\be
f_{k}^\mrm{SH} = \f{g_k/2}{\Delta_{{R}}+\w_k},
\ee
for the bosonic modes, where 
\be
\Delta_{{R}} = \Delta \la +f^\mrm{SH}| -f^\mrm{SH}\ra = \Delta e^{-2
\sum_{k>0}(f_{k}^\mrm{SH})^2} 
\label{DeltaR}
\ee
is the renormalized tunneling energy, which is thus
determined self consistently. The SH displacement naturally
interpolates between the zero tunneling case ($\Delta_R=0$) and the
zero dissipation limit ($g_k=0$).

Previous studies have shown that the SH state works relatively well for the
ground state at small coupling strength, $\alpha \lesssim 0.3$, and in the
scaling limit of $\Delta/\wc \ll 1$, but presents dramatic deficiencies
otherwise.~\cite{Chin_SubohmSBM11,Nazir,mccutcheon11,lee12,chin06} Its relative
success comes from the prediction of polaronic states, which minimize the
classical response of the environment. However, the modes entering each
spin-projected component in Eq.~(\ref{eq:SHansatz}) are fully uncorrelated, and
this misses a crucial aspect 
of the physics at
play.~\cite{Bera} One example of the failure of the SH state is the vanishing of
spin coherence at strong coupling, $\la \Psi_\mrm{SH}| \sigma_x|\Psi_\mrm{SH}\ra
= \Delta_R/\Delta = (\Delta e/\w_c)^{\alpha/(1-\alpha)}\to0$ for $\alpha\to1$,
while the exact Bethe Ansatz solution predicts a finite value $\la \sigma_x \ra
= \Delta/\wc$ in the scaling limit.~\cite{LeHurReview}

The nature of the missing correlations in the SH state was elucidated in our
recent publication~\cite{Bera} where we showed the importance of
\emph{antipolaronic} displacements at low energy, by which we mean displacements
$f_k$ of opposite value to $f_{k}^\mrm{SH}$ that are stabilized by quantum
tunneling effects.  In the following subsections, we present the details of this
extension --- the multi-polaron theory of the dissipative two-state system ---
that corrects all pathologies associated with SH theory.

\subsection{Building the multi-polaron ansatz: Two-polaron ground state}\label{sec:Ansatz}

\begin{figure}[tb]
	\centering
\includegraphics[width=0.9\columnwidth]{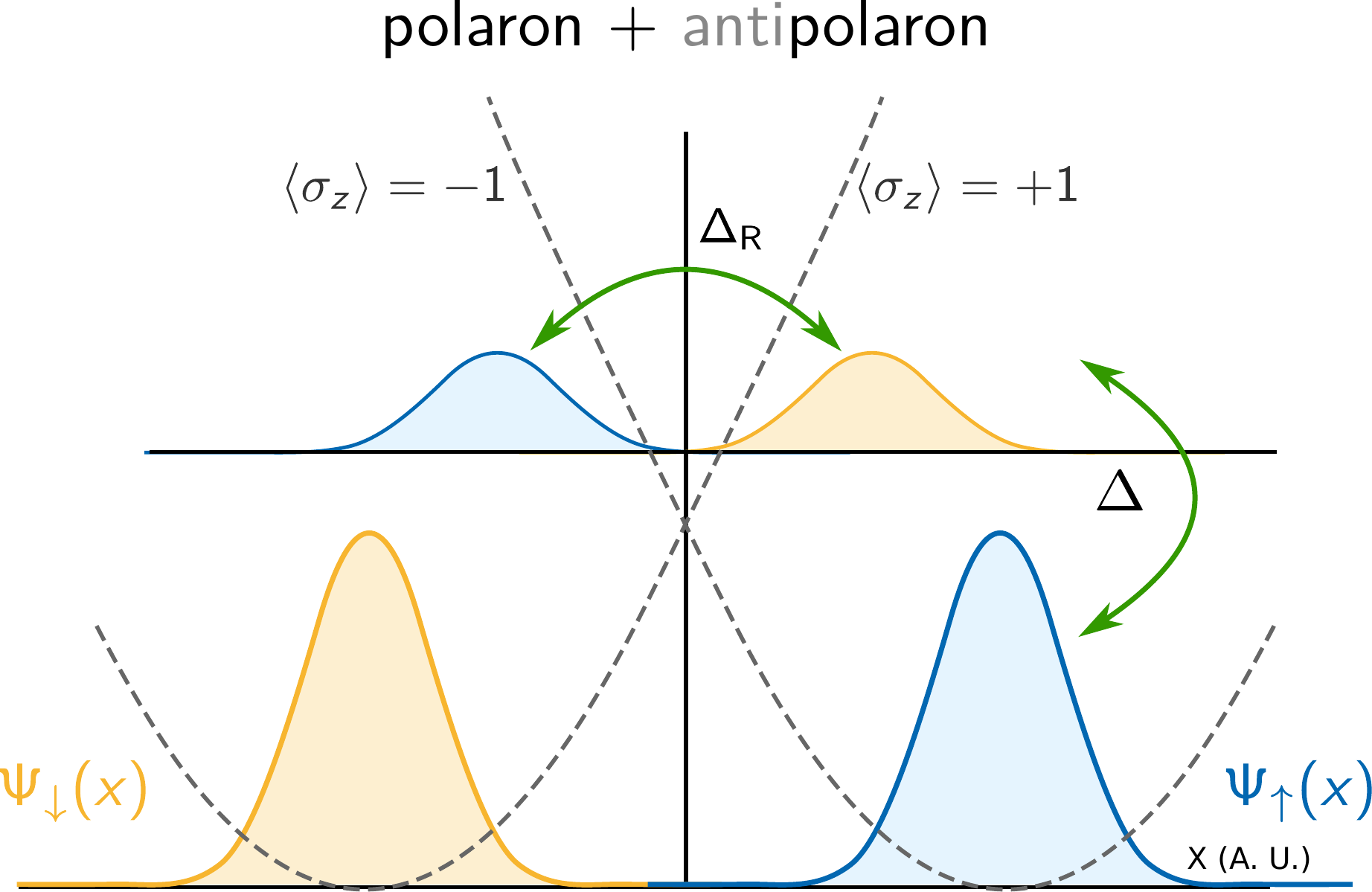}
\caption{Intuitive physical picture behind polaron and antipolaron
formation. Here, the wavefunction for a single oscillator mode is shown (blue
and gold curves correspond to the $\uparrow$ and $\downarrow$ projections, respectively).
The main weight of the wavefunction is carried by the polaronic components
(the two large lower lobes), but their overlap is exponentially small,
proportional to $\Delta_R$ as defined in Eq.~(\ref{DeltaR}) (horizontal green arrow). 
An enhanced tunneling energy is achieved through the emergence
of reduced weight antipolarons (the two small upper lobes), with an energy gain
proportional to the bare scale $\Delta$ (vertical green arrow). 
}
\label{fig:pol}
\end{figure}

We begin by explaining the physics at the heart of antipolaron formation, which
is rooted in the competition between tunneling (driven by $\Delta$) and
dissipation (due to the coupling to the bath).
For high frequency modes of the bath, $\w_k \gg \Delta$, the elastic energy dominates 
({i.e.} the energy associated to the displacement of the oscillators) and the 
displacement is given by the bare polaronic value $f_k = g_k/(2\w_k)$ for
mode $k$.
Thus, the renormalized spin tunneling energy is reduced by a factor roughly 
of order $\Delta_R/\Delta=e^{-\f{1}{2}\sum_{\w_k>\Delta}(g_k/\w_k)^2} \ll 1 $. 
Consequently, the spin-projected wavefunctions overlap poorly and the spin coherence 
is destroyed (see Fig.~\ref{fig:pol}). To overcome the resulting loss of tunneling energy, SH 
theory tends to adjust to smaller displacements for low energy modes, but 
will still predict an incorrect suppression of spin coherence at strong dissipation.
A possible way to maintain optimal tunneling energy, without sacrificing too much
elastic energy, is to allow quantum superposition of the classical polaronic component of
the wavefunction with coherent states that have negative displacements, which we have dubbed
antipolarons~\cite{Bera}, as sketched in Fig.~\ref{fig:pol}.

As an initial step towards building up the full multi-polaron ansatz, these qualitative ideas can be illustrated quite explicitly using a 
two-polaron variational ground state ansatz:
\begin{eqnarray}\label{variationalgs}
\ket{\Psi_{\rm 2pol}}&=&\ket{\uparrow}\left[C_1\ket{+f^{(1)}}
+C_2\ket{+f^{(2)}}\right]
\nonumber \\
&-&\ket{\downarrow}\left[C_1\ket{-f^{(1)}}
+C_2\ket{-f^{(2)}}\right],
\end{eqnarray}
where the bosonic part of the wavefunction again involves coherent states of the form
\begin{equation}
\ket{\pm f}=e^{\pm\sum_{k>0} f_k(b_k^{\dagger}-b_k)}\ket{0},
\end{equation}
defined as products of displaced states, where $\ket{0}$ represents all oscillators 
being in the vacuum state. The presence of a $\mathbb{Z}_2$ symmetry, namely invariance under 
($|\uparrow\big>\to|\downarrow\big>$, $|\downarrow\big>\to|\uparrow\big>$, $b_k\to-b_k$), 
and the need for minimizing the spin tunneling energy, enforces the chosen relative sign
between the up and down components of the ground state wavefunction in
Eq.~(\ref{variationalgs}).
Both functions $f_k^{(1)}$ and $f_k^{(2)}$ are
taken as free parameters, and will be varied to minimize the total ground state energy
$E_{\rm 2pol}=\bra{\Psi_{\rm 2pol}}H\ket{\Psi_{\rm 2pol}}$. In contrast to the usual
SH state (for which $C_2=0$),~\cite{Silbey1,Silbey2,Chin_SubohmSBM11,Nazir} this more flexible 
ansatz allows for the possibility of a superposition of variationally determined displaced 
oscillator states to be associated with each spin projection.
Normalization of $\ket{\Psi_{\rm 2pol}}$ implies the condition
\begin{equation}\label{normcondition}
1=2C_1^2+2C_2^2+4C_1C_2 \big<f^{(1)}|f^{(2)}\big>,
\end{equation}
with the usual form for the overlap of two coherent states, 
$\big<f^{(1)}|f^{(2)}\big> = 
e^{-\frac{1}{2}\sum_{k>0}(f_k^{(1)}-f_k^{(2)})^2}$. 
The two-polaron variational ground state energy is then given by
\begin{eqnarray}\label{evargs}
E_{\rm 2pol}&=&\bra{\Psi_{\rm 2pol}}\mc{H}\ket{\Psi_{\rm 2pol}}\nonumber\\
&=&-\Delta\Big[C_1^2\big<f^{(1)}|-\!f^{(1)}\big>
+C_2^2\big<f^{(2)}|-\!f^{(2)}\big>\nonumber\\ 
&&\;\;\;\;\;\;\;\:+2C_1C_2\big<f^{(1)}|\!-f^{(2)}\big>\Big]\nonumber\\
&&\:+2\sum_{k>0}\omega_k\Big[C_1^2(f_k^{(1)})^2+C_2^2(f_k^{(2)})^2
\nonumber\\
&&\;\;\;\;\;\;\;\;\;\;\;\;\;\;\;\;\;\;+2C_1C_2f_k^{(1)}f_k^{(2)}
\big<f^{(1)}|f^{(2)}\big> \Big]\nonumber\\
&&\:{-}2\sum_{k>0}g_k\Big[C_1^2f_k^{(1)}+C_2^2f_k^{(2)}
\nonumber\\
&&\;\;\;\;\;\;\;\;\;\;\;\;\;\;\;\;\;\;+C_1C_2(f_k^{(1)}+f_k^{(2)})\big<f^{(1)}|\!f^{(2)}\big> \Big].
\end{eqnarray}
In the limit that $C_2\rightarrow0$ (and so $C_1\rightarrow1/\sqrt{2}$) we recover the 
SH ground state energy~(\ref{eq:EnSh}). 

Returning to the two-polaron state of Eq.~(\ref{variationalgs}), we find that the ground 
state coherence is given by
\begin{eqnarray}\label{gscoh}
\langle\sigma_x\rangle&{}={}&-2\Big(C_1^2e^{-2\sum_{k>0}(f_k^{(1)})^2}
+C_2^2e^{-2\sum_{k>0}(f_k^{(2)})^2}\nonumber\\
&&\;\;\;\;\;\;\;{+}2C_1C_2e^{-\frac{1}{2}\sum_{k>0}(f_k^{(1)}+f_k^{(2)})^2}\Big),
\end{eqnarray}
while the magnetization $\langle\sigma_z\rangle=0$ by symmetry in the absence of
a magnetic field along $\sigma_z$ (unless one enters the polarized phase at $\alpha>1$ in 
the Ohmic spin-boson model). The two sets of displacements $f_k^{(1)}$ and $f_k^{(2)}$ are
variationally determined from the total energy $E_{\rm 2pol}$ of Eq.~(\ref{evargs}) according 
to $\partial E_{\rm 2pol}/\partial f_k^{(1)}=0$ and $\partial E_{\rm 2pol}/\partial f_k^{(2)}=0$, 
which gives the closed forms
\begin{equation}
\label{Supfkpol}
\resizebox{\hsize}{!}{$
f_k^{(1)}=
\frac{\frac{g_k}{2}\left[A_1 (C_2^2 \w_k + \Delta_2)-
A_2[C_1C_2\omega_k\big<f^{(1)}|f^{(2)}\big>+\Delta_{12}]\right]}
{(C_1^2\w_k+\Delta_1)(C_2^2\w_k+\Delta_2)-
[C_1C_2\omega_k\big<f^{(1)}|f^{(2)}\big>+\Delta_{12}]^2},
$}
\end{equation}
and
\begin{equation}
\resizebox{\hsize}{!}{$
f_k^{(2)}=
\frac{\frac{g_k}{2}\left[A_2 (C_1^2 \w_k + \Delta_1)-
A_1[C_1C_2\omega_k\big<f^{(1)}|f^{(2)}\big>+\Delta_{12}]\right]}
{(C_1^2\w_k+\Delta_1)(C_2^2\w_k+\Delta_2)-
[C_1C_2\omega_k\big<f^{(1)}|f^{(2)}\big>+\Delta_{12}]^2},
$}
\end{equation}
expressions which are valid for an arbitrary number of oscillator modes.
Hence, the generic $k$-dependence of the displacement is fully constrained by the
variational principle, which leaves a small set of effective parameters to
be determined self-consistently 
according to
\begin{eqnarray}
\Delta_1 &=& \Delta C_1^2 \big<-f^{(1)}|f^{(1)}\big>
+ \frac{\Delta}{2} C_1C_2 \big<-f^{(1)}|f^{(2)}\big>\nonumber\\
&&\:{+} C_1C_2 (-\tilde{\w}+\tilde{g}) \big<f^{(1)}|f^{(2)}\big>,\\
\Delta_2 &=& \Delta C_2^2 \big<-f^{(2)}|f^{(2)}\big>
+ \frac{\Delta}{2} C_1C_2 \big<-f^{(1)}|f^{(2)}\big>\nonumber\\
&&\:{+} C_1C_2 (-\tilde{\w}+\tilde{g}) \big<f^{(1)}|f^{(2)}\big>,\\
\Delta_{12} &=& 
\frac{\Delta}{2} C_1C_2 \big<-f^{(1)}|f^{(2)}\big>
+ C_1C_2 (\tilde{\w}-\tilde{g}) \big<f^{(1)}|f^{(2)}\big>,\nonumber\\\\
A_{1} &=& C_1^2+ C_1C_2 \big<f^{(1)}|f^{(2)}\big>,\\
A_{2} &=& C_2^2+ C_1C_2 \big<f^{(1)}|f^{(2)}\big>,\\
\tilde{\w} &=& \sum_k \w_k f_k^{(1)}f_k^{(2)},\\
\tilde{g} &=& \frac{1}{2} \sum_k g_k [f_k^{(1)}+f_k^{(2)}].
\end{eqnarray}
%
The one-polaron SH displacement
$f_k^\mathrm{SH}=0.5g_k/\big[\w_k+\Delta\big<-f^{(1)}|f^{(1)}\big>\big]$ 
can be recovered from Eq.~(\ref{Supfkpol}) by letting $C_2=0$.
One can also check that $f_k^{(2)} \simeq f_k^{(1)}$ for
$\omega_k\to\infty$, while $f_k^{(2)}\simeq -f_k^{(1)}$ for
$\omega_k\to0$ in the limit of strong dissipation. Thus, the antipolaron displacement satisfies 
the expected adiabatic to non-adiabatic crossover as a function of energy $\w_k$, and
this physical picture is naturally incorporated into the variational theory.

\begin{figure}[t]
	\centering
\includegraphics[width=0.9\columnwidth]{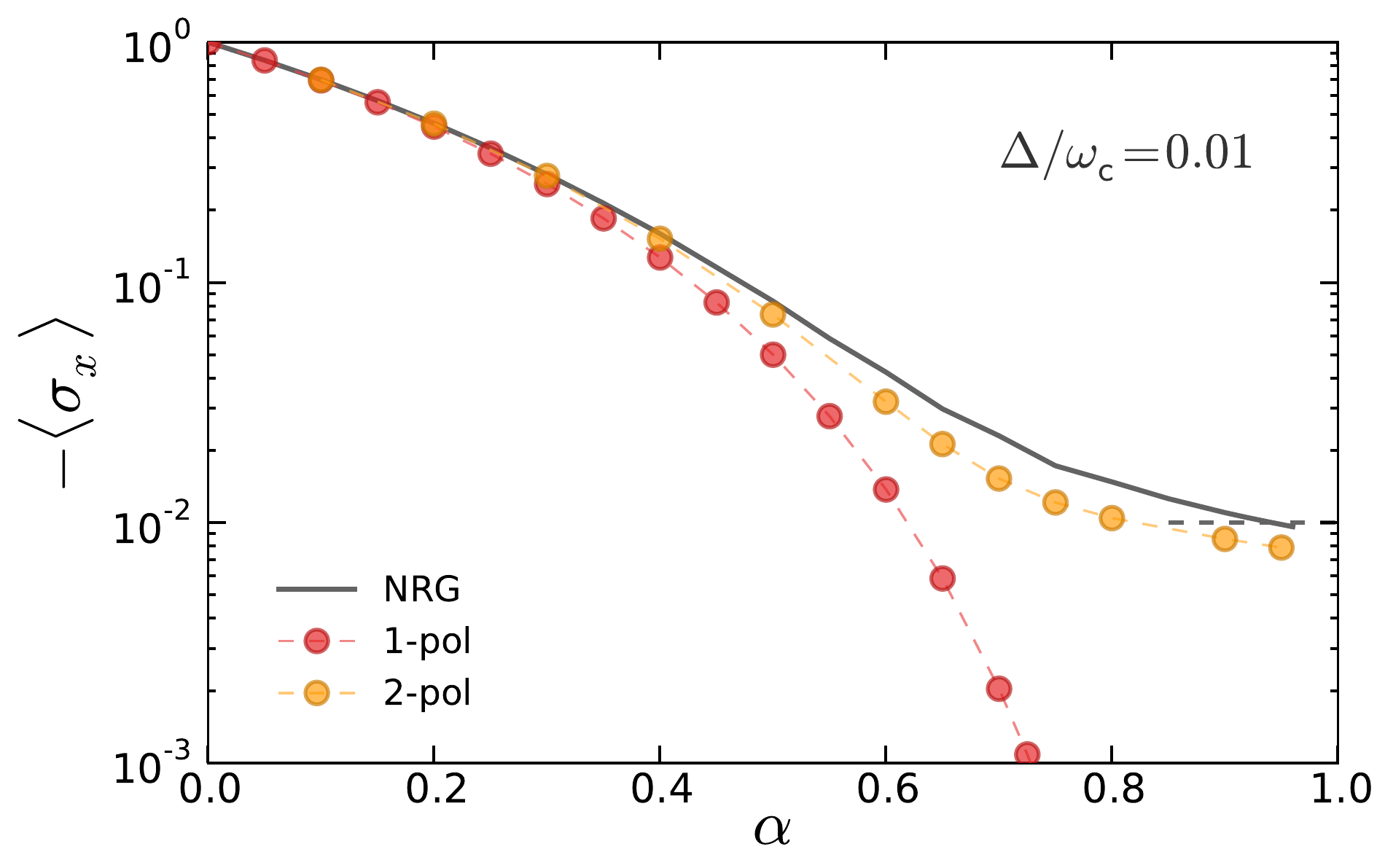}
\caption{
Ground state spin coherence in the Ohmic spin boson model, $\la\spin{x}\ra$, as
a function of dissipation strength $\alpha$ for tunneling energy
$\Delta/\wc=0.01$. The results of NRG calculations are shown as a solid line and
the dots denote predictions from the single-polaron SH state (1-pol) and the
two-polaron ansatz (2-pol). The SH state clearly fails to capture the correct
spin coherence at large $\alpha$. The black dashed line on the right denotes
the expected Bethe-Ansatz value at the transition point $\alpha=1$.
} \label{fig:Sx2Pols}
\end{figure}

In Fig.~\ref{fig:Sx2Pols} we show how the addition of a second polaron
dramatically improves the ground state spin coherence at strong dissipation in
the two-polaron ansatz. In particular, by incorporating polaron-antipolaron
overlap, the two-polaron ansatz correctly predicts an enhancement of the spin
coherence at large $\alpha$ in comparison to the one-polaron SH state, which
instead incorrectly predicts a strong suppression given by the vanishing scale
$\Delta_R/\Delta$ defined in Eq.~(\ref{DeltaR}). We see that the spin coherence derived from the two-polaron
ansatz is already in very good qualitative agreement with converged NRG results,
though there are some quantitative differences, such as an underestimation of
$|\la\spin{x}\ra|$ as $\alpha\rightarrow1$. This motivates the inclusion of
additional polarons into the ground state ansatz, which should act to further
enhance polaron-antipolaron overlap, and thus bring the coherence into full
agreement with the NRG results.


\subsection{Multi-polaron ground state expansion}\label{sec:expansion}
The considerations of the previous subsection lead us to propose a generalized
multi-polaron ground state ansatz that captures the complete structure of the
entanglement 
built into the bath:
\be
|\Psi_{\rm GS}\ra =  \sum_{n=1}^{N_\mrm{pols}}  C_n \Big[ |+ \fn{n} \ra
\otimes \Su - |-\fn{n} \ra \otimes \Sd \Big],
\label{eq:GS}
\ee
where $|\pm \fn{n} \ra$ once more denote multi-mode coherent states, 
$
|\pm \fn{n} \ra = e^{\pm
\sum_{k>0} \fkn \Le(b_k^\dg-b_k\Ri)} |0\ra
$,  
with $\fkn$ now the displacement of mode $k$ for the $n^\mrm{th}$ variational
coherent state. Here, $C_n$ characterizes the weight of the different coherent
state components within the ground state wavefunction and 
$N= \la \Psi_{\rm GS} | \Psi_{\rm GS} \ra $ denotes the norm.
In the limit
$N_\mrm{pols} \rightarrow \infty $ the above ground state wavefunction allows an
arbitrary superposition of polaron and antipolaron states, which in turn
should capture all the environmental correlations that are missing in 
SH variational theory.
In the opposite limit, i.e.~$N_{\mathrm{pols}}=1$, Eq.~\eqref{eq:GS}
reduces to standard single-polaron SH theory. 

It is important to stress here that the set of coherent states required to
achieve convergence is not necessarily very large, as we shall see later on. 
Because of the energetic requirements of the problem, the allowed displacements 
are strongly constrained, which consequently reduces the total number of
coherent states needed in the expansion of Eq.~\eqref{eq:GS}.  For instance,
the high energy oscillator modes $\w_k \gg \Delta$ typically fall onto
the bare displacement value $f_k^{(n)} \simeq g_k/(2\w_k)$.
Also, spin tunneling energy can be gained whenever the displacements become opposite
in sign to this classical value, which forces the coherent states to cross
over from positive to negative displacements, and thus alternate between
polaronic and antipolaronic branches.
Finally, in the low energy limit, $\w_k\to 0$, all displacements tend to vanish in
order to preserve spin coherence. We shall see below that all of these constraints
are well obeyed by the numerical solution of the multi-polaron equations.

\subsection{Solving the multi-polaron equations}
We now present details of the procedure used to determine the displacements
$f_k^{(n)}$ and weights $C_n$ entering the multi-polaron ansatz~(\ref{eq:GS}).
First, we need to compute the energy of this variational ground state 
$E_\mrm{GS}= \la \Psi_{\rm GS} | \mc{H}
|\Psi_{\rm{GS}}\ra / \la \Psi_\mrm{GS} | \Psi_\mrm{GS}\ra $  within the spin-boson
model [Eq.~\eqref{eq:SB}]:
\begin{eqnarray}
\label{eq:Menergy}
E_\mrm{GS} &=& 
-\f{\Delta}{N} \sum_{n,m}^{N_\mrm{pols}} C_n C_m \la \fn{n} | -\fn{m} \ra \\
\nonumber
&& + \f{2}{N} \sum_{n,m}^{N_\mrm{pols}} C_n C_m \la \fn{n} | \fn{m} \ra
\sum_{k>0} \w_k \fkn \fkm  \\ 
\nonumber
&&- \f{1}{N} \sum_{n,m}^{N_\mrm{pols}} C_n C_m  \la \fn{n} | \fn{m} \ra \sum_{k>0} 
g_k (\fkn + \fkm),
\end{eqnarray}
with the norm $N= 2 \sum_{n,m}^{N_\mrm{pols}} C_n C_m \la \fn{n} | \fn{m} \ra$. 
The overlaps of different coherent states are given by
$
\la \fn{n} | \pm \fn{m} \ra = e^{-\f{1}{2}\sum_{k>0} (\fkn \mp \fkm)^2}
$.  
The first term in $E_{\rm GS}$ describes the spin tunneling contribution,
while the last two terms contain the displacement energy of the oscillators.
Eq.~(\ref{eq:Menergy}) clearly reduces to the SH 
energy, Eq.~\eqref{eq:EnSh}, in the limit $N_\mrm {pols} = 1$.
All observables are determined once the variational parameters $f_k^{(n)}$ and $C_n$
are known. For instance, the spin coherence reads simply:
\be
\la \spin{x} \ra = -\f{\sum_{n,m}^{N_\mrm{pols}} C_n C_m e^{-\f{1}{2}\sum_{k>0}(\fkn + \fkm)^2
} }{ \sum_{n,m}^{N_\mrm{pols}} C_n C_m e^{-\f{1}{2}\sum_{k>0}(\fkn - \fkm)^2} }.
\label{eq:spinavg}
\ee
From the above expression for the spin coherence, it is possible to understand the
failure of the SH ansatz. The spin coherence in SH theory is
given by the expression
$\la \spin{x} \ra =e^{-2 \sum_{k>0} (f^{\mrm pol.}_k)^2}  =  \Delta_{
R}/\Delta$, and is thus determined solely by the renormalized tunneling amplitude. While
a vanishing (Kondo) energy scale is indeed expected for $\alpha\to1$, the spin coherence
should remain finite.
In contrast, the multi-polaron ansatz contains additional contributions to
the spin coherence, with pre-factors $e^{-\f{1}{2}\sum_{k>0}(\fkn+\fkm)^2}$. 
Antipolarons, namely $\fkn\simeq -g_k/(2\w_k)$, will tend to cancel the polaronic
displacement $\fkm\simeq g_k/(2\w_k)$, thus
helping to stabilize the spin coherence even at large dissipation. 
In other words, the correlations captured by the multi-polaron ansatz [Eq.~\eqref{eq:GS}] 
strongly enhance the spin coherence by introducing entanglement within the bath.   

We have devised an algorithm to solve for the displacements efficiently. 
After truncation of the spectral density to a number
$N_\mrm{modes}$ of modes (which could be on either a linear or logarithmic grid depending
on the regime considered), we face the determination of
$N_\mrm{pols}\times (N_\mrm{modes}+1)$ unknown displacements and weights, that
should solve the set of coupled non-linear equations $\partial E_\mrm{GS}/\partial
\fkn = 0$ and $\partial E_\mrm{GS}/\partial C_n = 0$.
With increasing numbers of polarons and modes this method becomes 
impractical
because of the large number of self-consistent parameters, and numerical
instabilities typically associated with root-finding routines.
Instead, we use a direct minimization of the full energy functional, but
even so, finding the global minima of $E_\mrm{GS}$ is no small task.
However, dramatic computational gains can be achieved by exploiting the energetic
requirements on the displacements $f_k^{(n)}$ discussed in
Sec.~\ref{sec:expansion}. 
We thus use a two step minimization procedure, where the first step consists of
a global minimization routine using displacements that are parametrized
by a small set of unknown energy scales $\{\Omega_i^{(n)}\}$:
\be
\fkn = \f{-g_k/2}{\w_k + \Delta_{R}} \prod_{i=1}^{I^{(n)}} \f{\w_k -
\Omega^{(n)}_i}{\w_k + \Omega^{(n)}_i}.
\label{eq:fkParam}
\ee
The rationale behind this expression is that it permits each displacement to
cross-over from polaronic to anti-polaronic branches whenever the mode frequency $\w_k$
crosses a node of the displacement $\fkn$ at energy $\w_k=\Omega^{(n)}_i$. For a given multi-mode coherent
state $|f^{(n)}\ra$ we allow an arbitrary number $I^{(n)}$ of nodes in the routine, 
but typically the energy minimization will favour states with one node over states
with two nodes (and so on).
Using the restricted form of $\fkn$ given in Eq.~(\ref{eq:fkParam}) we perform a global 
stochastic optimization of the energy using standard simulated annealing
techniques~\cite{Corana87, Goffe94}, which allows us to fix the parameters
$\Delta_{R}$, $\Omega^{(n)}_i$ and $C_n$, and already gives an excellent
variational approximation to the exact ground state. The convergence can be
further improved in a second minimization step, by implementing a full variational 
determination of the $N_\mrm{pols}\times (N_\mrm{modes}+1)$ unknown displacements and 
weights. This is performed using the final result of the global parametrized
solution as an input, which is fed into either to an efficient conjugate
gradient program~\cite{Hager05, Hager06, Hager06Survey, Hager13} or to a limited
memory BFGS method routine~\cite{Byrd95, Zhu97} depending on the parameter
range. Here, all displacements and weights are varied simultaneously.
In order to facilitate comparison to NRG
results (see Appendix~\ref{AppendixA} for details on the NRG simulations) and to limit the number of modes in 
the strong dissipation regime, the optimization is performed on a logarithmic discretization 
of the bath. We shall now present and discuss the results obtained by using this 
algorithm.

\subsection{Results and Discussion}
\begin{figure}[tb]
	\centering
\includegraphics[width=0.9\columnwidth]{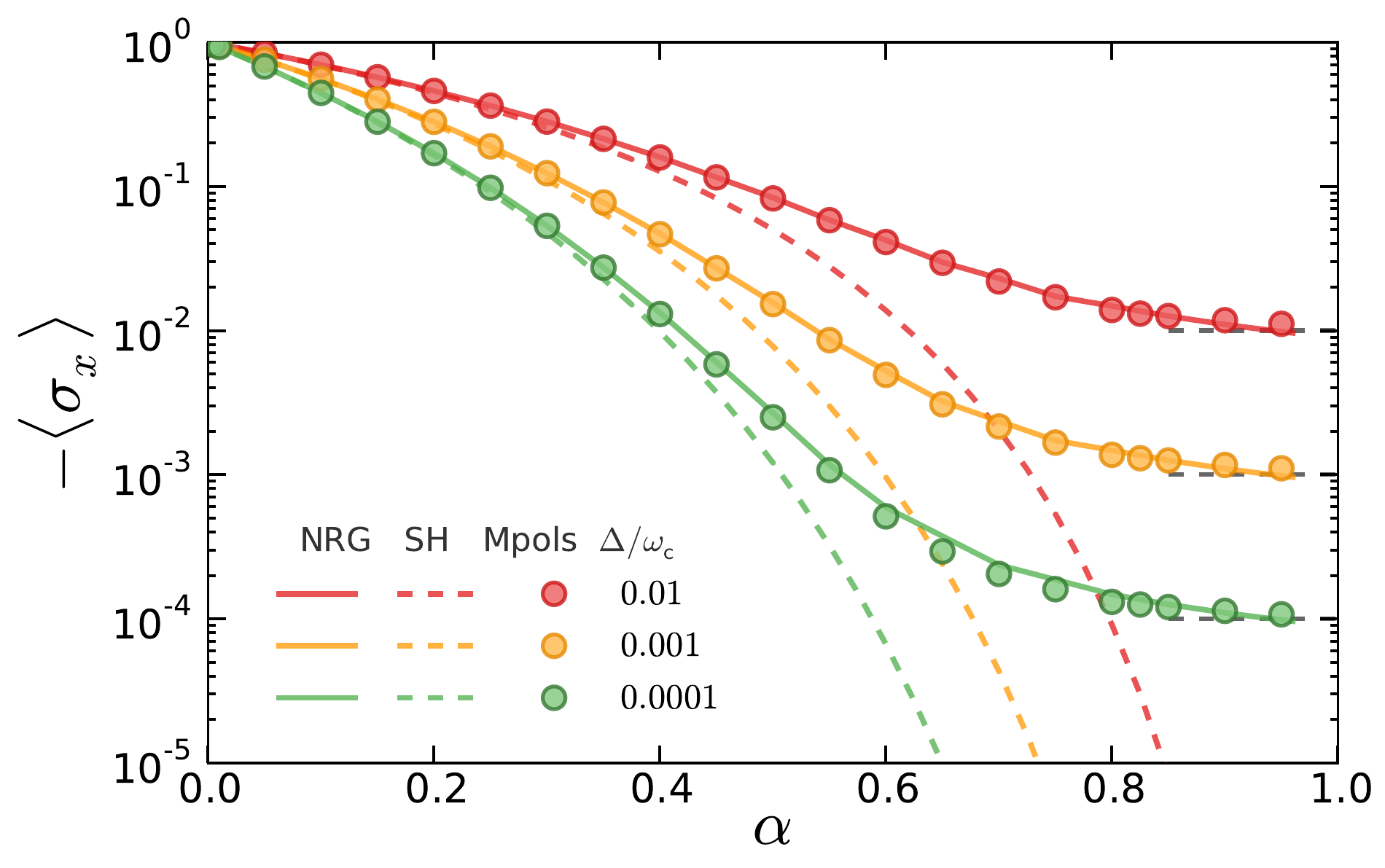}
\caption{
Spin coherence in the Ohmic spin boson model, $\la\spin{x}\ra$, as a function of
dissipation strength $\alpha$ for three different values of the tunneling energy,
$\Delta/\wc=0.0001,0.001,0.01$ (bottom to top). The solid lines show results
of NRG calculations, and the dots 
(labeled ``Mpols'') mark the converged results from our multi-polaron
ansatz with $N_\mrm{pols}=6$ polaronic states. Again, we see that the SH 
results~(dashed lines) clearly fail to capture the spin coherence at large 
$\alpha$.}
\label{fig:Sx}
\end{figure}
\begin{figure*}[tb]
	\centering
\includegraphics[width=0.99\linewidth]{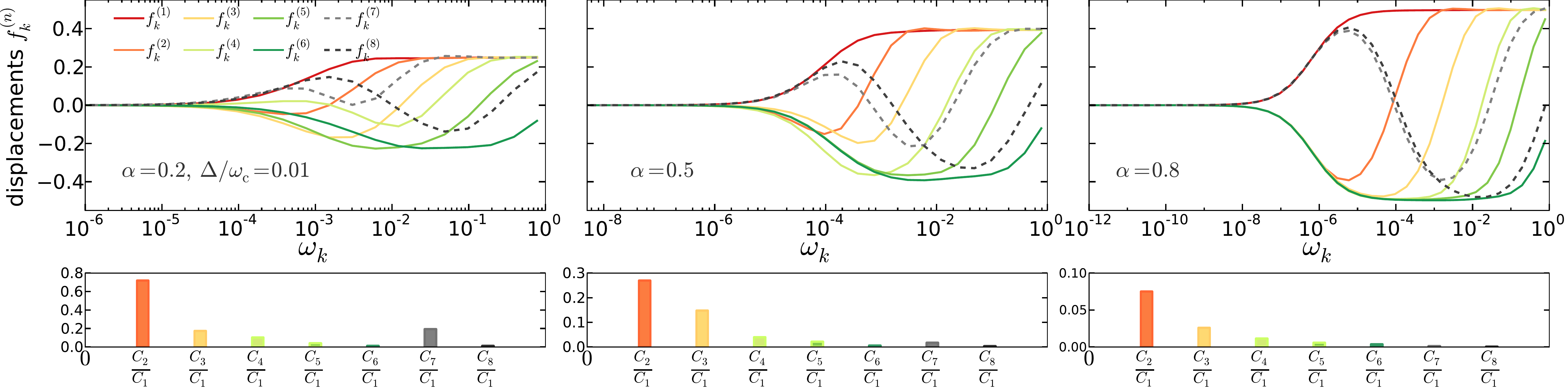}
\caption{
Variationally determined displacements $\fkn$ (top panels) and
weights $C_n$ (bottom panels) for three different values of dissipation
$\alpha=0.2,0.5,0.8$ (left to right panels), calculated from the multi-polaron 
variational ansatz [Eq.~\eqref{eq:GS}] using $N_\mrm{pols}=8$ coherent states.  
These calculations are performed for $\Delta/\wc = 0.01$ and on a logarithmic 
grid with $\Lambda = 2$ (see Appendix~\ref{AppendixA}). All displacements tend to cluster to the
same positive values at high frequencies (polarons), and cross over to negative
values in a complicated fashion (antipolarons). We note the presence of one state
with no node $f^{(1)}$ (the fully polaronic state given by the solid red line), five 
states with one node $f^{(2)}\ldots f^{(6)}$ (other solid lines), and two states 
with two nodes $f^{(7)}$ and $f^{(8)}$ (dashed lines).
The lower three panels show the weights $C_n/C_1$, normalized to
the weight of the fully polaronic state, corresponding to the above
oscillator displacements~(using the same color code).}
\label{fig:fkAlpha}
\end{figure*}
\begin{figure}[tb]
\includegraphics[width=0.98\columnwidth]{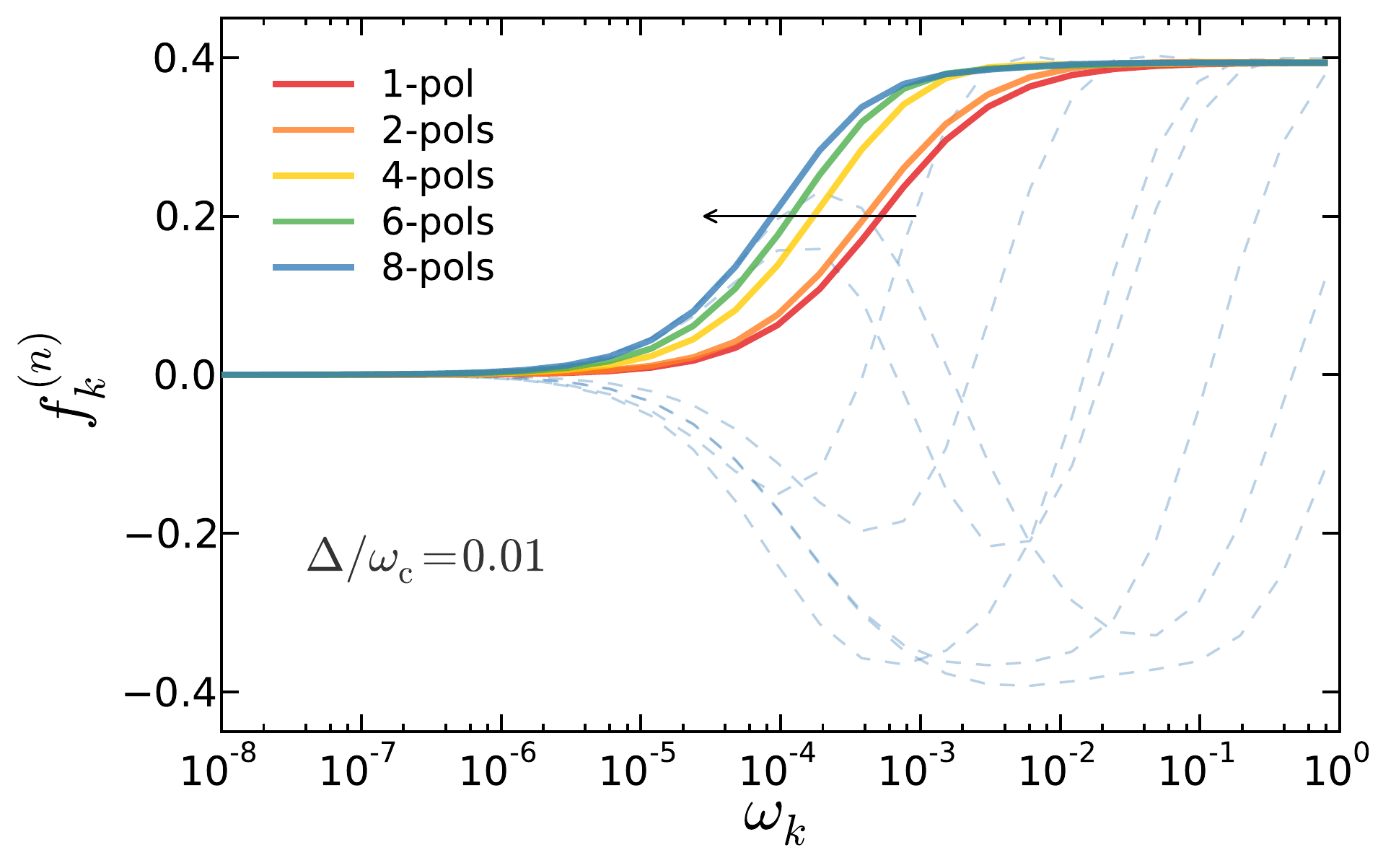}
\caption{Polaronic (zero-node) displacement $f_k^{(1)}$ computed for
an increasing number of coherent states, $N_\mrm{pols}=1,2,4,6,8$ (full curves, from right to left) in 
Eq.~(\ref{eq:GS}). Parameters: $\Delta/\w_c=0.01$ and $\alpha=0.5$.
The multi-node displacements for $N_\mrm{pols}=8$ are shown as dotted
curves.}
\label{fig:Fk1Npols}
\end{figure}
\begin{figure*}[tb]
	\centering
\includegraphics[width=0.98\linewidth]{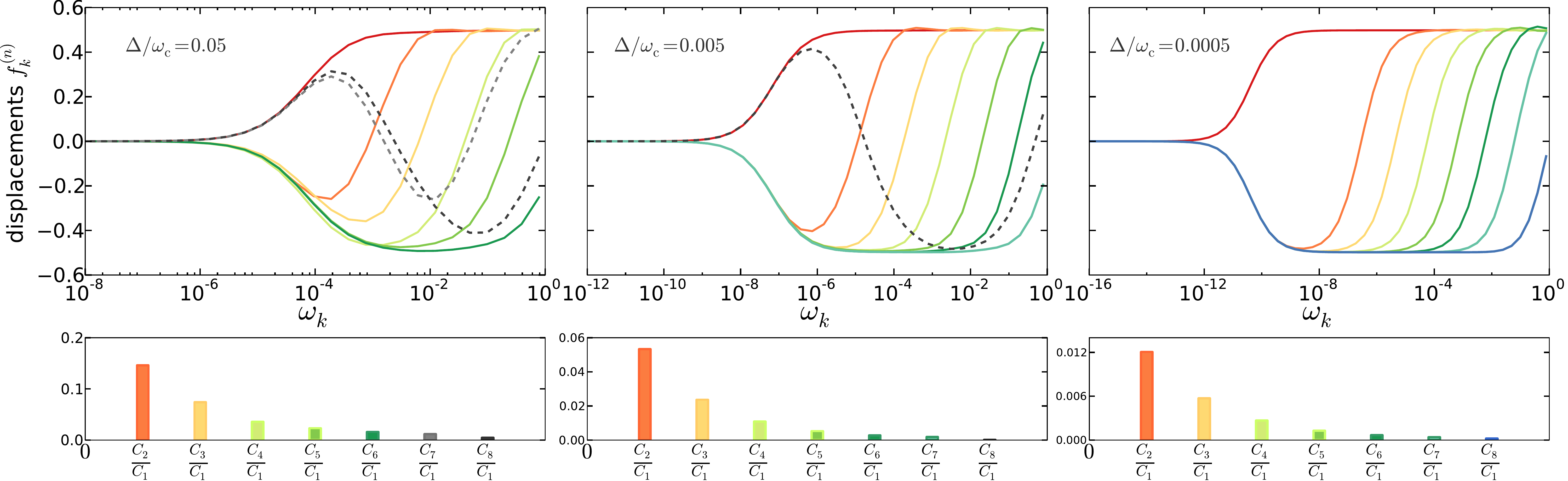}
\caption{Displacements $\fkn$ and weights $C_n$ as in
Fig.~\ref{fig:fkAlpha}, but for fixed dissipation $\alpha=0.8$ and decreasing
values of the tunneling amplitude, $\Delta/\wc=\{0.05, 0.005, 0.0005\}$ (left to
right).}
\label{fig:fkDelta}
\end{figure*}
\begin{figure}[tb]
	\centering
\includegraphics[width=0.99\columnwidth]{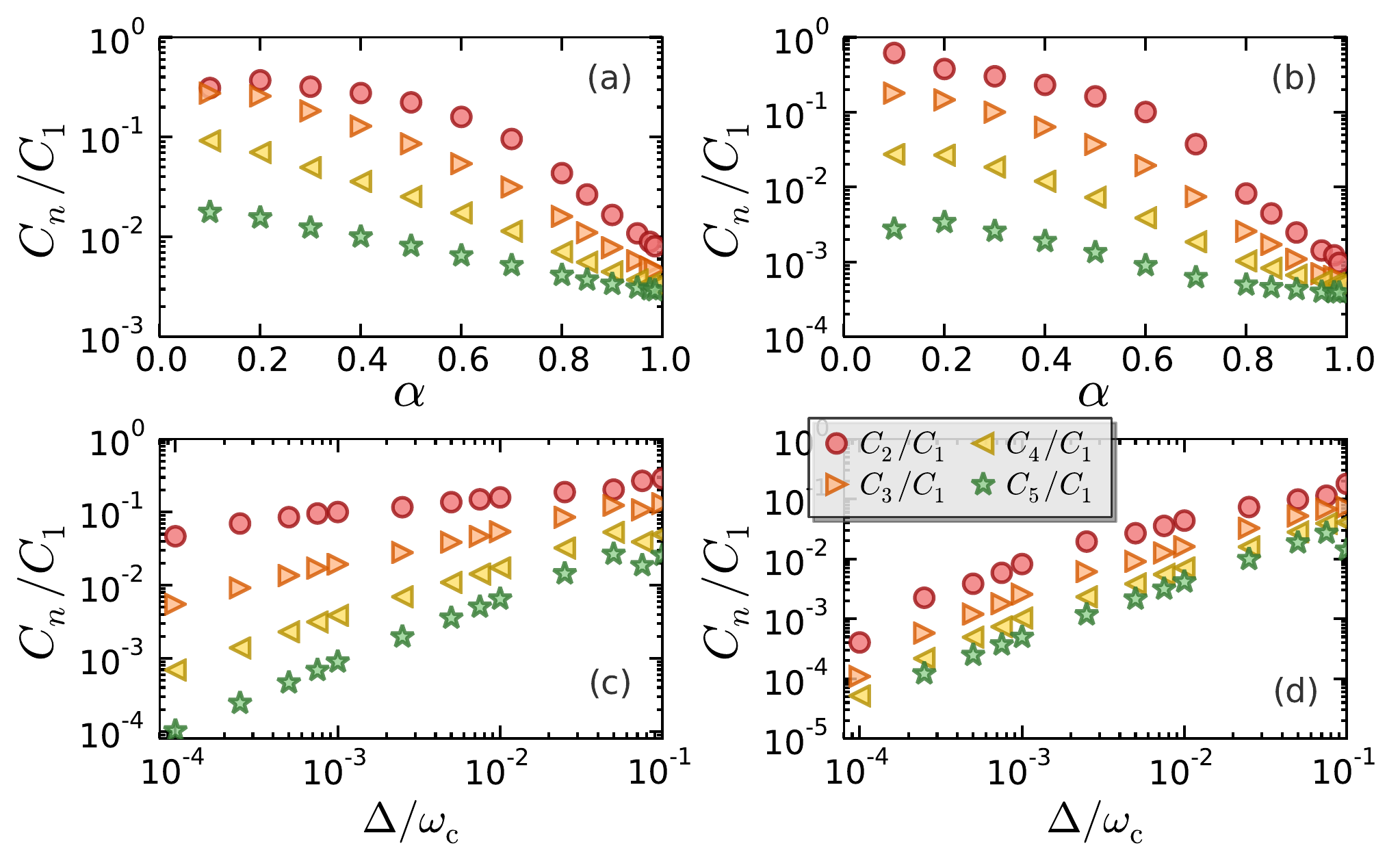}
\caption{Weight $C_n/C_1$ of the multi-mode coherent state $|f^{(n)}\ra$ ($n>1$),
normalized to the weight of the main polaron component $|f^{(1)}\ra$, as a
function of spin-bath coupling $\alpha$~(upper panels) and tunneling amplitude
$\Delta$~(lower panels). These calculations were performed for $N_\mrm{pols}=5$
coherent states, and $\Delta/\wc=0.01$ in panel {\bf (a)}, $\Delta/\wc=0.001$ 
in panel {\bf (b)}, $\alpha=0.6$ in panel {\bf (c)}, and $\alpha=0.8$ in panel {\bf (d)}.
The weights saturate to finite values in the limit $\alpha\to1$ (top panels),
highlighting the non-trivial nature of the wavefunction at the quantum critical
point, while the weights vanish in the $\Delta\to0$ bare polaron limit (bottom
panels).}
\label{fig:pVsAlDel}
\end{figure}
\subsubsection{Spin coherence}
In Fig.~\ref{fig:Sx} we show the ground state spin coherence  $\la\spin{x}\ra$ as a function
of spin-bath coupling (dissipation) strength $\alpha$ for three different values of
$\Delta/\wc = \{0.01, 0.001, 0.0001\}$, calculated using the multi-polaron
ansatz in Eq.~\eqref{eq:GS} with only $N_{\rm pols}=6$ coherent states. For comparison, the SH results~(dashed
lines) are also shown along with results from the NRG calculations~(solid lines). As mentioned
previously, SH theory completely fails to capture the spin coherence 
for any value of $\Delta/\wc$ at large coupling ($\alpha \gtrsim 0.3$), and incorrectly 
predicts an exponentially small renormalized spin tunneling energy 
$\la\spin{x}\ra = \Delta_{R}/\Delta = (\Delta e/\wc)^{\alpha/(1-\alpha)}$. 
In contrast, the extra environmental correlations
coming from different antipolaron overlaps $
e^{-\f{1}{2}\sum_{k>0}(\fkn+\fkm)^2} $ (for $n \ne m$) in the multi-polaron ansatz preserve the spin
coherence, as seen by  the excellent agreement with the NRG simulations.
Finally, close to the quantum critical point $\alpha \approx 1$, the saturation 
of spin coherence to the exact Bethe Ansatz value $\sim \Delta/\wc$ is also captured 
properly by our proposed multi-polaron wavefunction. It is also worth stressing
the rapid convergence of the multi-polaron ansatz with only a small
number of coherent states for all values of $\Delta/\wc$ and $\alpha$ shown here. 
For the spin coherence, this is shown explicitly in Fig.~2 of 
our previous publication \cite{Bera} and is discussed below for 
other quantities. 

\subsubsection{Displacements.} 
In analyzing the displacements we shall start by considering the role of dissipation on the entanglement structure of
the ground state wavefunction of the spin-boson model.
Fig.~\ref{fig:fkAlpha} shows a set of ground state bosonic displacements $\fkn$
for three different values of the dissipation strength $\alpha$ calculated using the multi-polaron ansatz of
Eq.~\eqref{eq:GS}. Here we present the solution obtained with $N_\mrm{pols} = 8$ 
coherent states, to show the presence of displacements with up to two nodes.
The zero node displacement, noted $f_k^{(1)}$, is completely analogous to
the SH polaronic displacement and encodes the main classical response
of the bath.
The one node displacements, denoted $f_k^{(2)}\ldots f_k^{(6)}$ in
Fig.~\ref{fig:fkAlpha}, nicely illustrate the crossover from polaron behavior for
high energy modes to antipolaronic behavior for lower energy modes.
We observe in Fig.~\ref{fig:Fk1Npols} that the main displacement $f_k^{(1)}$ in the multipolaron
case differs from the single polaron results quantitatively, and typically
we find that the exact renormalized scale $\Delta_R^{\mrm{exact}}$ is 
significantly smaller than the SH value 
$\Delta_R = \Delta (\Delta e/\wc)^{\alpha/(1-\alpha)}$ defined in
Eq.~(\ref{DeltaR}). Here
$\Delta_R^{\mrm{exact}}$ corresponds physically to the scale where all
displacements vanish in the converged multipolaron ground state 
(see Fig.~\ref{fig:Fk1Npols}), which can be formally defined 
as follow. One can compute the average ground state displacement for
the spin-up component of the full ground state wavefunction:
$f_k^{\mrm{ground}} = \la \Psi_{\rm GS} | 
(\ak + \akd) (1+\sigma_z)/2|
\Psi_{\rm GS} \ra $.
The overlap between positively and negatively shifted effective polarons with
displacements $f_k^{\mrm{ground}}$ provides then a precise estimate of the
true renormalized tunneling energy:
$\Delta_R^{\mrm{exact}} = \Delta e^{-2\sum_k(f_k^{\mrm{ground}})^2}$.
In fact, this strong renormalization from $\Delta_R$ to $\Delta_R^{\mrm{exact}}$ 
comes from the proliferation of antipolarons in the energy range
$[\Delta_R,\w_c]$, which pushes the scale $\Delta_R$ downwards, as shown in 
Fig.~\ref{fig:Fk1Npols}.

More surprising is the emergence of states with two nodes in the
displacements
$f_k^{(7)}$ and $f_k^{(8)}$, as allowed in our starting guess
[Eq.~(\ref{eq:fkParam})]. 
Such states are thus polaronic at high energy, antipolaronic at intermediate energy, 
and polaronic again at low energy.
The appearance of these states can be understood as follows: because coherent
states are not orthogonal to each other, it is not useful to add up a very large
number of one-node states with different crossover frequencies $\Omega_1^{(n)}$,
as the overlap between any pair of such one-node states will tend to one. In
fact, it instead becomes favorable to create states that gain some combination
of potential and tunneling energy, {i.e.} states that can approach  as close as
possible the polaron and the antipolaron branches,  while being as different as
possible from the one-node states. Having displacements with two nodes is an
obvious way to cope with these constraints, and we can envision at this stage
that the complete expansion of the wavefunction can be rationalized in terms of
displacements with an ever increasing number of nodes. 
We also note from Fig.~\ref{fig:fkAlpha} that such two node state may have
a larger weight than the one node displacements (for instance $C_7$ in the
bottom left panel), which would naively imply that the correction brought
by $f_k^{(7)}$ is larger than the ones obtained at the previous orders. This
is not so actually, because the various polaron states in the wavefunction 
decomposition are not orthogonal to each other. Thus, a state with substantial 
overlap with the main displacement $f_k^{(1)}$(as is actually the case with 
the two-node state $f_k^{(7)}$) is actually allowed to have a large weight, although 
it provides in the end a small correction to the actual complete wavefunction.


We emphasize that the weights associated with multi-node coherent states
are rapidly suppressed for $n>4$ (see bottom panels in Fig.~\ref{fig:fkAlpha}),
demonstrating the rapid convergence of our ansatz. Another observation is that
the fully optimized displacements become quantitatively very close to the trial
form of Eq.~\eqref{eq:fkParam} at large dissipation, underlining the advantage
of the chosen initial parametrization.  We believe, although we have not yet
proved, that such a simplification for the form of the displacements is related
to the emergent universal scaling properties that are inherent in the underlying
Kondo physics~\cite{LeHurReview,Leggett,Guinea85, EmeryLutherPRB} for
$\alpha\simeq1$.  Our multi-polaron
state thus provides interesting insights into the nature of entanglement within
the Kondo cloud, taking advantage of the simplifications brought about by the
natural emergence of coherent states in the bosonic language appropriate here.
Such a precise understanding is to our knowledge still lacking for the fermionic
Kondo model, owing to the complexity in parametrizing particle-hole excitations
in a Fermi gas beyond the reach of perturbation theory.

We now consider the role of tunneling energy on the microscopic nature of
the many-body wavefunction.
Fig.~\ref{fig:fkDelta} shows the $k$-dependence of the oscillator displacements 
and the related weights for three values of $\Delta/\wc=\{0.05, 0.005, 0.0005\}$ 
at $\alpha=0.8$, again for a total of $N_\mrm{pols}=8$ multi-mode coherent states.
Similarities, as well as interesting differences, appear as compared to the effect
of increasing dissipation. We observe that the renormalized tunneling $\Delta_R\propto
\Delta^{1/(1-\alpha)}$ drops as expected by increasing $\alpha$ (Fig.~\ref{fig:fkAlpha}) or 
by decreasing $\Delta$ (Fig.~\ref{fig:fkDelta}), as seen by the behavior of
the characteristic energy at which all displacements vanish together (note the difference in frequency scales from the left to right upper plots in these figures).
For the chosen parameters in Fig.~\ref{fig:fkDelta}, 
the two-node displacements become unfavorable as compared to new 
one-node displacements at decreasing $\Delta$.
We thus find that the scalings $\alpha\to1$ and $\Delta\to0$ are {\it not}
equivalent, because the two-node displacements are stable in the former.
In addition, one expects that the $\Delta=0$ limit should be trivial as it amounts simply to 
bare polarons, while the $\alpha\to1$ limit is associated with a Kosterlitz-Thouless quantum 
phase transition.
The corresponding behavior of the weights, see Fig.~\ref{fig:pVsAlDel}, does indeed
support this view: all $C_n$ ($n>1$) saturate to a finite value for $\alpha\to1$,
emphasizing the inherently strongly entangled nature of the wavefunction, while
all $C_n$ ($n>1$) rapidly drop to zero for $\Delta\to 0$, recovering the expected
bare polaron limit.

\subsection{Spectroscopy of entanglement entropy}
\label{sec:entropy}

\begin{figure}[tb]
        \centering
\includegraphics[width=0.9\columnwidth]{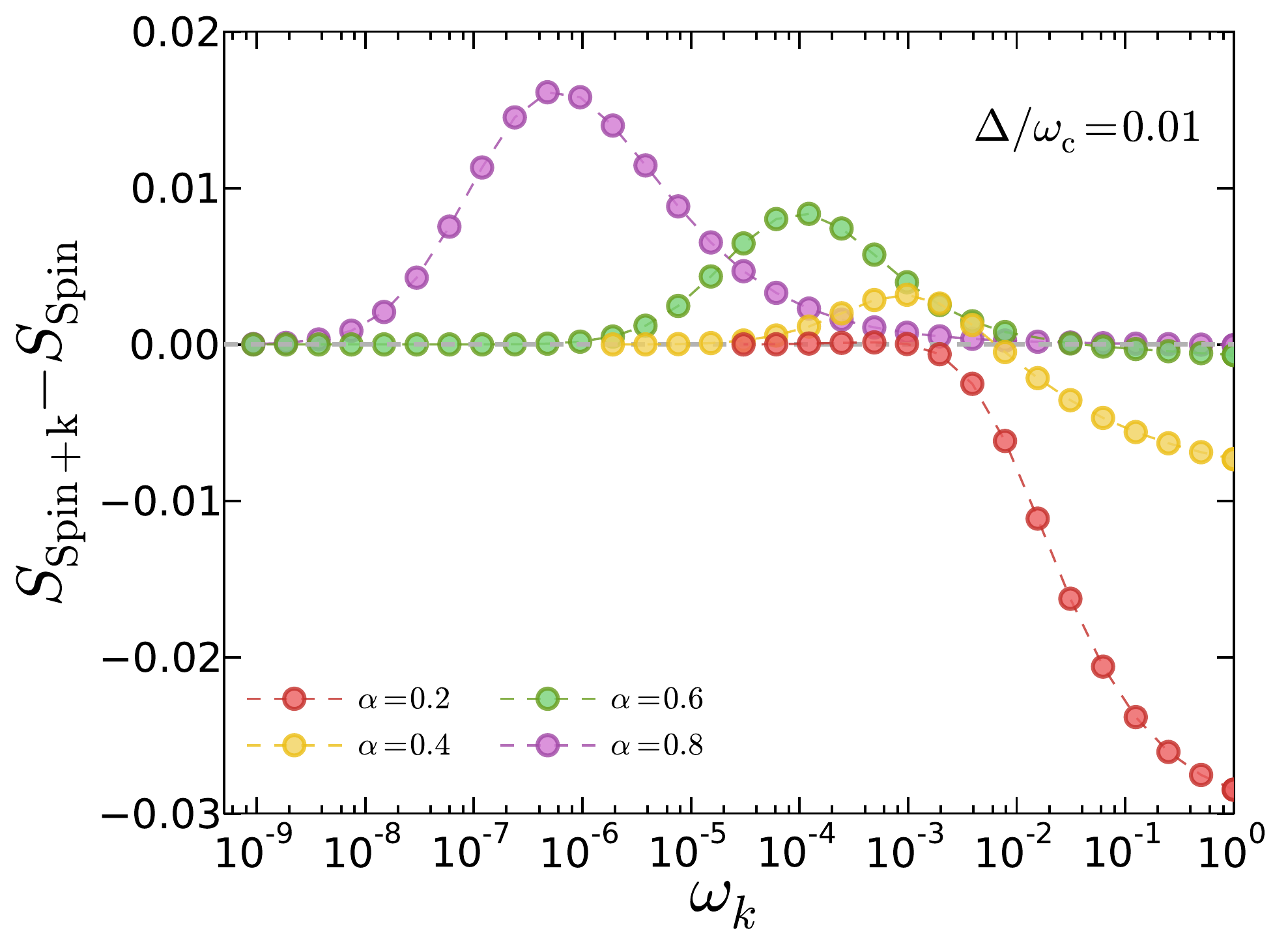}
\includegraphics[width=0.9\columnwidth]{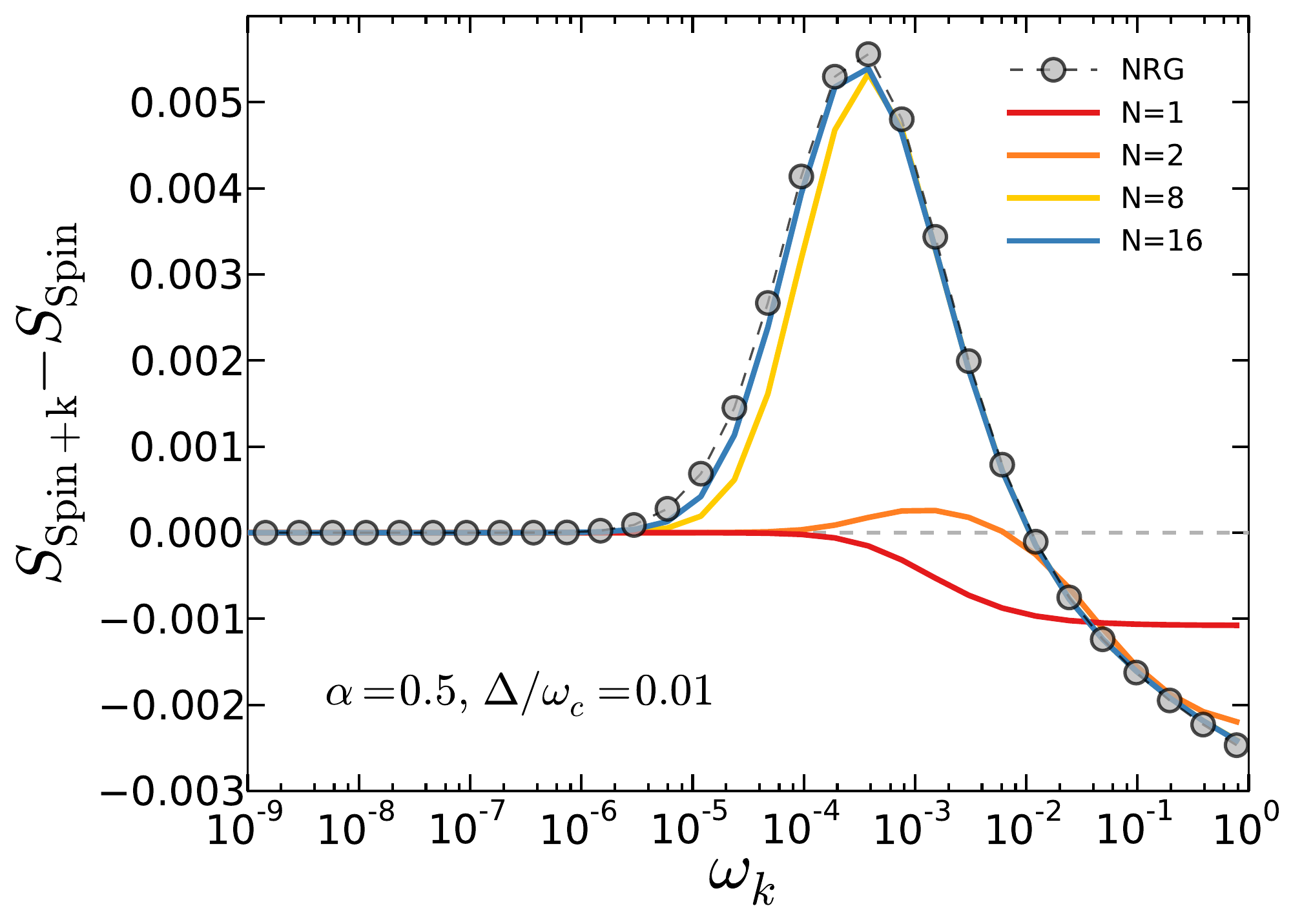}
\caption{
Excess entanglement $\rho^2$-entropy of the subsystem composed of the
qubit with a given oscillator $k$-mode. The top figure displays the NRG results for
four values of dissipation
$\alpha=0.2,0.4,0.6,0.8$, showing a qualitative change of behavior for
$\alpha>0.5$ (see text). The bottom panel, computed for $\alpha=0.5$ within the
NRG and the coherent state expansion~(\ref{eq:GS})
for increasing coherent state content ($N_\mrm{pols}=1,2,8,16$), shows that the large
and positive entropy excess is built from the entanglement generated by
multi-polaronic components. }
\label{fig:Rh2}
\end{figure}

We now wish to assess the ground state entanglement among the environmental
states, suggested by the coherent state expansion~(\ref{eq:GS}), more directly.
For this purpose, we define two reduced density matrices from which an ``excess
entropy'' measure will be constructed. First, the reduced density matrix of the
spin alone is obtained from the pure ground state density matrix
$\rho_{\mathrm{tot}}=\big|\Psi\big>\big<\Psi\big|$ by tracing out all of the
bosonic environmental modes, \begin{equation} \label{rhospin}
\rho_{\mathrm{spin}} = \mathrm{Tr}_{\mathrm{env}} \big|\Psi\big>\big<\Psi\big|.
\end{equation} The second reduced density matrix is obtained by tracing out all
modes except the qubit degree of freedom together with an arbitrary bath  mode
with given quantum number $k$; this defines a spin and $k$-mode excluded
environment denoted ``$\mathrm{env/spin}+k$''. The reduced ground state density
matrix in the joint qubit and $k$-mode subspace reads \begin{equation}
\label{rhospink} \rho_{\mathrm{spin}+k} = \mathrm{Tr}_{\mathrm{env/spin}+k}
|\Psi\big>\big<\Psi|.  \end{equation} From these reduced density matrices,
corresponding entanglement entropies can be defined. We choose to work with the
$\rho^2$-entropy since it can be readily evaluated for the multi-polaron ansatz
wavefunction: $S_\mathrm{spin}= 1-\mathrm{Tr} [\rho_\mathrm{spin}^2]$ is the
entanglement entropy of the spin with the bath, and $ S_{\mathrm{spin+k}} =
1-\mathrm{Tr} [\rho_{\mathrm{spin}+k}^2]$ is the entanglement entropy of the
spin plus $k$-mode subsystem with the other environmental modes. To assess the
entanglement within the bath, consider the difference between these two
entropies, $ S_{\mathrm{spin+k}}-S_{\mathrm{spin}}$, the excess entropy due to
the entanglement of the mode $k$ with the rest of the environment. This quantity
is plotted in Fig.~\ref{fig:Rh2} from both the NRG data and the multi-polaron
ansatz (see Appendix~\ref{AppendixB} for details of the calculation) and reveals
most sensitively the nature of the many-body ground state.


The excess entropy is mostly negative for small dissipation, $\alpha<0.5$ (see
the NRG calculations in the top panel of Fig.~\ref{fig:Rh2}), as expected from
the correlations characterizing the SH ansatz~(\ref{eq:SHansatz}), which is
built solely from non-entangled environmental states within each spin-projected
component.  In contrast, at strong dissipation $\alpha>0.5$, the excess entropy
becomes positive and shows a strikingly large enhancement near the scale
$\Delta_R~{\mathrm{exact}}$, which we interpret as being due to entanglement within the bath of
oscillators.  This interpretation is confirmed by the direct comparison with the
entropy computed from the coherent state expansion~(\ref{eq:GS}), see the bottom
panel of Fig.~\ref{fig:Rh2}. Indeed, the entropy peak gradually builds up as
coherent states are added into the wavefunction thus generating 
additional environmental entanglement.  Note especially the large energy window
where the entropy peak develops: the excess entanglement spreads from low to
high frequency modes.  The existence of inter-mode bosonic correlations on a
wide energy range also makes an interesting connection to the underlying
(although hidden in the spin-boson model) fermionic Kondo
physics.~\cite{Leggett,LeHurReview} 

This excess entropy is thus a very sensitive measure of the subtle non-classical
correlations among the bath modes that are generated by their coupling to the
qubit.

\subsection{Convergence properties}

\begin{figure}[tb]
        \centering
\includegraphics[width=0.99\columnwidth]{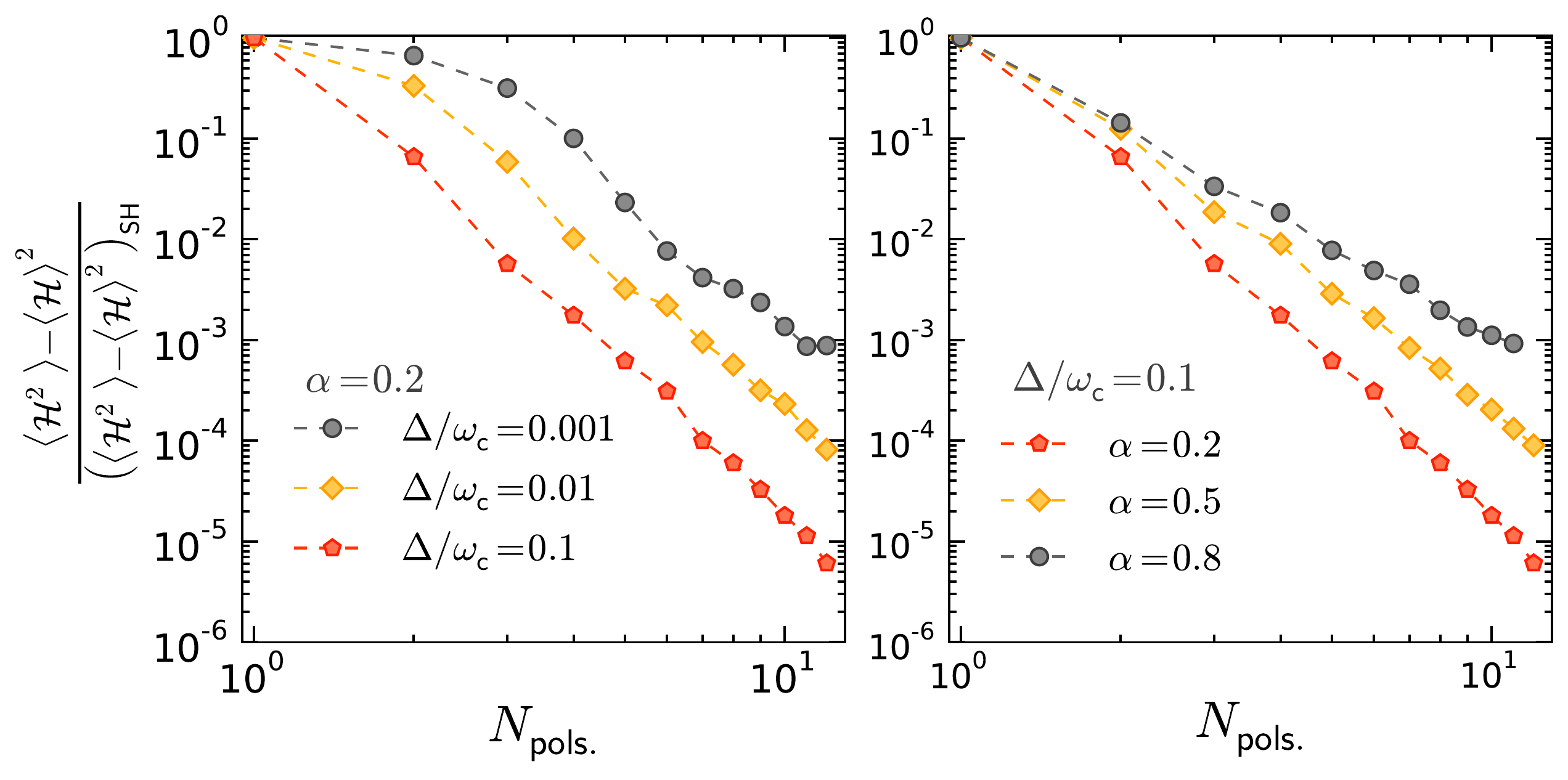}
\caption{Energy variance $\big<\mc{H}^2\big>-\big<\mc{H}\big>^2$ 
(normalized to the
SH result) as a function of the number $N_\mrm{pols}$ of coherent
states in the expansion~(\ref{eq:GS}) for several values of the tunneling rate
(left panel) and different dissipation strengths (right panel).}
\label{fig:H2}
\end{figure}

In the last part of this section, we address the convergence properties of our
coherent state expansion~(\ref{eq:GS}). While the variational principle and
the overcompleteness property of the coherent state basis suggest
convergence to the full many-body ground state of the spin boson model, we would
like to check that the numerical procedure used to determine the unknown
parameters $(f_k^{(n)},C_n)$ is not detrimental to the correct convergence. For
this purpose, we compute the energy variance
$\big<\mc{H}^2\big>-\big<\mc{H}\big>^2$, which can be expressed analytically
from the trial state~(\ref{eq:GS}), see Appendix~\ref{AppendixC}, and is
displayed in Fig.~\ref{fig:H2}. 

We find that the energy variance vanishes quickly for a large number of
polarons, typically in a power-law fashion, with an exponent that depends on the
dissipation strength. This shows that an exact eigenstate is approached by
increasing the number of coherent states in the trial wavefunction, 
suggesting rapid convergence of the expansion. Such efficiency of our
algorithm is rooted in two important aspects of the polaron theory: first,
the Silbey-Harris polaron state captures the majority of the full many-body
wavefunction, so that only perturbative corrections have to be obtained
from the antipolaron states; second, the energetic constraints discussed
previously reduce the phase space of available displacements constituting
the corrections to the Silbey-Harris state, which guarantees fast
convergence of the polaron expansion.

\section{Biased Spin-Boson model}\label{sec:biased}
\subsection{Hamiltonian and motivation}
In the second part of the paper, we extend our unbiased multi-polaron ansatz 
[Eq.~\eqref{eq:GS}] to incorporate the effect of a bias $\epsilon$ within the 
spin-boson model: 
\be
\mc{H}_\mrm{bias} = \f{\epsilon}{2}\spin{z} + \f{\Delta}{2} \sigma_x  -
\f{\sigma_z}{2}\sum_{k > 0} g_k (b_k + b_k^{\dg}) + \sum_{k>0} \w_k
b_k^{\dag} b_k.
\label{eq:SbBias}
\ee
We again consider an Ohmic spectral density, and we shall focus primarily on
the dependence on $\epsilon$ of various observables, namely the ground state spin coherence
$\la \sigma_x \ra$ and population difference (magnetization) $\la \sigma_z \ra$, which
now becomes non-zero even below $\alpha=1$. 
In terms of experimental realization, in the context of
superconducting circuits, for example, $\Delta$ corresponds to the Josephson energy that couples a charge qubit 
to an array of superconducting
junctions,~\cite{LeHurJosephson,GoldsteinJosephson,Wilhelm} while
$\epsilon$ is controlled by a local gate that
shifts the degeneracy of the two successive charge states.
On theoretical grounds, the inclusion of bias is interesting in several ways.
First, single-polaron SH theory does a very poor job at large dissipation
for biased systems, and shows strong artifacts for $\alpha > 1/2$,
for instance predicting an incorrect magnetization jump.~\cite{Nazir} Proposals to cope with some
of these defects have been made,~\cite{Nazir} but were justified on the basis of physical arguments
rather than from a clear mathematical procedure grounded in the variational
principle. For this reason they did not offer a fully optimized framework.

In the following we provide a generalization of the unbiased multi-polaron
ansatz to treat the ground state of the biased spin-boson
model, Eq.~\eqref{eq:SbBias}. We shall show that our biased multi-polaron 
ansatz not only corrects all pathologies associated with the SH ansatz, 
but also converges to exact results obtained from the Bethe Ansatz.

\subsection{Asymmetric Silbey-Harris theory}

Let us start by considering the generalization of the unbiased one-polaron state, Eq.~(\ref{eq:SHansatz}), to the biased case.
In principle we should now allow {\it a priori}
different displacements (no longer equal and opposite) and weights 
for the up and down components of the wavefunction, leading to a
spin-asymmetric SH trial state:
\be
|\Psi_{\rm aSH}^\mrm{bias}\ra =  \f{1}{\sqrt{p^2+q^2}} 
\Big[p | f_{\ua} \ra \otimes \Su - q |f_{\da} \ra \otimes \Sd \Big].
\label{eq:aSHbias}
\ee
The motivation for introducing this form of wavefunction is to be able to
capture both the $\alpha=0$ and $\Delta=0$ limits exactly, 
in contrast to the symmetric ansatz of Eq.~(\ref{eq:GS}), which is in fact also used in the
literature for the biased case.~\cite{Silbey1,Silbey2,Nazir,mccutcheon11} 
Applying the variational principle leads to simple closed-form expressions for the 
spin-dependent displacements:
\be\label{eq:fkaSH}
f_{\ua, \da} = 
	\f{g_k (\tilde{\epsilon} \mp \w_k)}{2\w_k(\chi + \w_k)},
\ee
with the parameters $\tilde{\epsilon}$ and $\chi$ (as well as the
weights $p$ and $q$) determined by a set of non-linear self-consistent
equations. Clearly, the parameter $\chi$ is connected to the renormalized
tunneling amplitude, but in contrast to the unbiased case, the displacements
no longer vanish in the $\w_k\to0$ limit, but rather approach the
renormalized value $[\tilde{\epsilon}/(2\chi)] g_k/\w_k$. This can be understood physically 
from the fact that finite positive bias, $\epsilon>0$,
enforces a finite negative value of $\la \sigma_z\ra < 0 $, which in turn 
provides a finite and positive displacement to the oscillators due to the linear
coupling term in the spin-boson Hamiltonian~(\ref{eq:SbBias}).
We stress that the oscillators associated with the up and down components of the wavefunction thus
displace in the same ``direction'' at low energy due to the bias, in contrast 
to a polaron-type displacement, which has opposite values for $f_\uparrow$ and
$f_\downarrow$,
as can be seen from the limit of large $\w_k$ in Eq.~(\ref{eq:fkaSH}).
 
\subsection{Biased multi-polaron ground state ansatz}
We shall now extend our multi-polaron ansatz, Eq.~\eqref{eq:GS}, to incorporate
the symmetry breaking features associated with asymmetrically displaced oscillator 
states:
\be
|\Psi_{\rm GS}^\mrm{bias}\ra =  \f{1}{\sqrt{N}}\sum_{n=1}^{N_\mrm{pols}} \Big[p_n |
\fn{n}_{\ua} \ra \otimes \Su - q_n |\fn{n}_{\da} \ra \otimes \Sd \Big].
\label{eq:GSbias}
\ee
Here, $|\fn{n}_{\ua, \da} \ra  =  e^{\sum_{k>0} \fa{n}{k\, (\ua, \da)} (\bkd
-\bk) } |0\ra$. 
In the limit $\fn{n}_{\ua} = -\fn{n}_{\da}$ and $p_n = q_n$, the above ansatz
reduces to unbiased multi-polaron ansatz of Eq.~\eqref{eq:GS}, and for
$N_\mrm{pols}=1$, we recover the asymmetric SH state of Eq.~(\ref{eq:aSHbias}).

As in the unbiased spin-boson model, the coherent state basis is over complete
and thus has ample flexibility to capture the main energetic constraints at
play in the presence of bias too. These are of three types: (i) the formation of polarons at high energy;
(ii) the formation of antipolarons at intermediate energies, with an increasingly
complex nodal structure to the displacements; and (iii) the saturation to a finite
displacement at vanishing energy controlled by the bias field.
We shall show below that these physical considerations completely characterize the
rich entanglement content of the wavefunction in the biased case. 
Again, we find that a reasonably small number of multi-mode coherent
states is sufficient for good convergence, comparing this time to exact Bethe Ansatz results.

\subsection{Solving the biased multi-polaron equations}
The variational multi-polaron energy for the biased spin-boson model, $E_\mrm{GS}^\mrm{bias}= \la\Psi_\mrm{GS}^\mrm{bias}|\mc{H}_\mrm{bias}|\Psi_\mrm{GS}^\mrm{bias}\ra /
\la\Psi_\mrm{GS}^\mrm{bias}|\Psi_\mrm{GS}^\mrm{bias}\ra $, is given by
\begin{widetext}
\begin{align}
	E_\mrm{GS}^\mrm{bias}=& \f{1}{N}\sum_{n,m}^{N_\mrm{pols}}\Bigg[ -\Delta p_n q_m \la \fa{n}{\ua}|
	\fa{m}{\da}\ra	+  p_n p_m \la \fa{n}{\ua}| \fa{m}{\ua}\ra
	\sum_{k>0} \w_k \fa{n}{k\,\ua}\, \fa{m}{k\,\ua} +
	 q_n q_m \la \fa{n}{\da}| \fa{m}{\da}\ra \sum_{k>0} \w_k \fa{n}{k\,\da}\, \fa{m}{k\,\da} \nn  \\
	&-  p_n p_m \la \fa{n}{\ua}| \fa{m}{\ua}\ra
	\sum_{k>0} \f{g_k}{2} \Le(\fa{n}{k\,\ua} +  \fa{m}{k\,\ua}\Ri) +
	q_n q_m \la \fa{n}{\da}| \fa{m}{\da}\ra
	\sum_{k>0} \f{g_k}{2} \Le(\fa{n}{k\,\da} +  \fa{m}{k\,\da}\Ri) \nn \\ 
	&+  \f{\epsilon}{2} \Le( p_n p_m \la \fa{n}{\ua}|
\fa{m}{\ua}\ra - q_n q_m \la \fa{n}{\da}| \fa{m}{\da}\ra \Ri) \Bigg],
\label{eq:MenergyBias}
\end{align}
\end{widetext}
where 
$
N = \sum_{n,m}^{N_\mrm{pols}} \Le(p_n p_m \la \fa{n}{\ua}| \fa{m}{\ua}\ra + q_n
q_m \la \fa{n}{\da}| \fa{m}{\da}\ra\Ri)
$
is the normalization of the complete ground state wavefunction and 
$
\la \fa{n}{\ua, \da} | \fa{m}{\ua, \da}\ra  =  e^{-\f{1}{2}\sum_{k>0} \Le[\fa{n}{k\,
\ua, \da}-\fa{m}{k\, \ua, \da}\Ri]^2}
$
is the overlap between different coherent states. 
It is straightforward to see that in the unbiased limit
$\epsilon = 0$, the ground state energy of Eq.~\eqref{eq:MenergyBias} reduces to
the unbiased spin-boson ground state energy of Eq.~\eqref{eq:Menergy}. 
Here, we perform a variation of the energy with respect to all free parameters
within the problem, namely the spin-polarized displacements $\fa{n}{\ua,\da}$, and
their related weights $p_n$ and $q_n $.

All observables are determined once these parameters are known.
For instance, the spin coherence $\la\spin{x}\ra$ and magnetization 
$\la\spin{z}\ra$ are given by the compact expressions:
\begin{align}
\label{eq:SxBias}
\la\spin{x}\ra &= \f{-
\sum_{n,m}^{N_\mrm{pols}} p_n q_m \la \fa{n}{\ua}|
	\fa{m}{\da}\ra
}
	{
		\sum_{n,m}^{N_\mrm{pols}} \Big[p_n p_m \la \fa{n}{\ua}| \fa{m}{\ua}\ra + q_n
		q_m \la \fa{n}{\da}| \fa{m}{\da}\ra\Big] },\\
\la\spin{z}\ra &= \f{
\sum_{n,m}^{N_\mrm{pols}} 
\Big[ p_n p_m \la \fa{n}{\ua}|
\fa{m}{\ua}\ra - q_n q_m \la \fa{n}{\da}| \fa{m}{\da}\ra \Big]
}
{\sum_{n,m}^{N_\mrm{pols}} \Big[ p_n p_m \la \fa{n}{\ua}| \fa{m}{\ua}\ra + q_n
q_m \la \fa{n}{\da}| \fa{m}{\da}\ra\Big]}.
\label{eq:SzBias}
\end{align}
In the absence of bias, $\fa{n}{\ua} = -\fa{n}{\da}$ and $p_n = q_n$, hence we
readily recover a vanishing magnetization, $\la\spin{z}\ra = 0$, from 
Eq.~\eqref{eq:SzBias}. 

Guided by the unbiased scenario, we perform a two step
minimization of the total energy, Eq.~\eqref{eq:MenergyBias}. 
Note that the number of free parameters to minimize here 
is doubled in comparison to the unbiased multi-polaron ansatz, so that finding a
reliable and fast algorithm is quite crucial.
Again, we first parametrize the displacements using only a small number 
of parameters to allow for an efficient global optimization: 
\be\label{eq:fkParamBias}
\fa{n}{\ua, \da} = 
	\f{g_k (\tilde{\epsilon} \pm \w_k)}{2\w_k (\chi + \w_k)}
\prod_{i=1}^{I^{(n)}}
	\f{\w_k - \Omega_{i (\ua,\da)}^{(n)}}{\w_k + \Omega_{i (\ua,\da)}^{(n)}},
\ee
where $\tilde{\epsilon},\, \chi$, and $\Omega_{i (\ua,\da)}^{(n)}$ are 
variational parameters, along with the weights $p_n,\, q_n$. Here,
$\Omega_{i (\ua,\da)}^{(n)}$ denotes a set of crossover frequencies for each coherent
state, with the index $i$ spanning the allowed nodes. 
In practice, we start the minimization procedure with the possibility of having 
an arbitrary nodal structure.
We also emphasize here that this parametrization goes beyond the ansatz proposed 
in Ref.~\onlinecite{Nazir}, which is in fact not variational. Although our parametrization is not yet {\it fully} variational
at the end of the first minimization stage, it 
captures the essential 
energetic constraints discussed previously.
In the second stage, we then take the output of the global minimization routine 
to initialize
the final local optimization routine, which performs a local 
search within the entire set of parameters (displacements and weights),
thereby completing the numerical variational minimization of the total energy in Eq.~(\ref{eq:MenergyBias}).

\subsection{Results and Discussion}
\subsubsection{Displacements.} 
\begin{figure}[tb]
	\centering
\includegraphics[width=0.95\columnwidth]{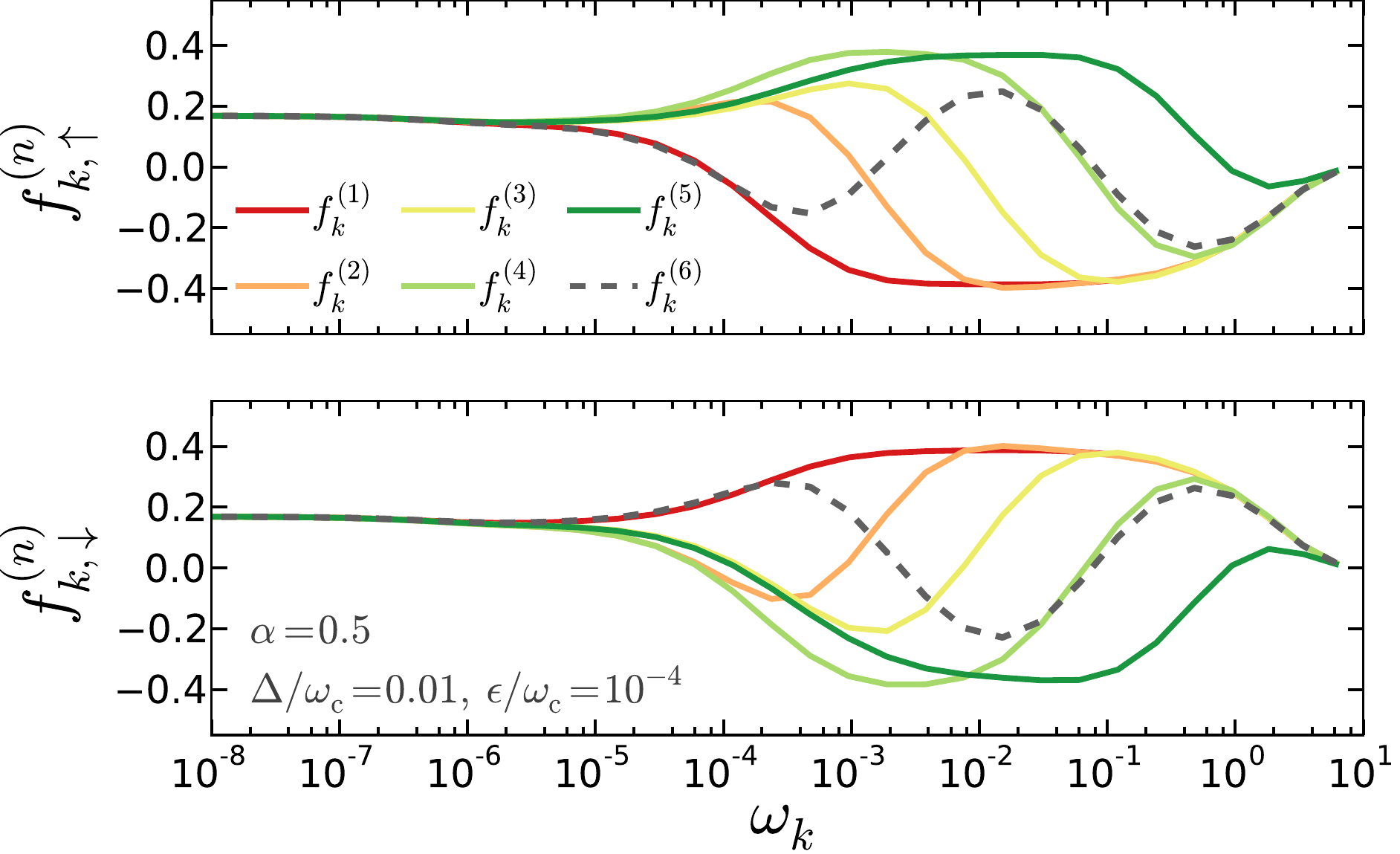}
\caption{Variational displacements $\fa{n}{k,\ua/\da}$ for the biased spin-boson
model, Eq.~\eqref{eq:SbBias}. To facilitate comparison with Bethe Ansatz
results, all multi-polaron calculations were performed using an exponential
high-energy cutoff in the spectral density.  Thus, the displacements fall off
exponentially at $\w_k>\w_c$ instead of being cut off sharply as in the previous
plots in the unbiased case (for instance in Fig.~\ref{fig:fkAlpha}).  We note
again the emergence of antipolaron displacements, but additional nodal structure
emerges due to the saturation of the displacements to a finite value at
$\w_k\to0$ (for instance $f_{k,\ua}^{(1)}$ shows a node although it corresponds
to the main polaronic component). Here, the parameters used are $N_\mrm{pols} =
6$, $\Delta/\wc = 0.01$, $\alpha = 0.5$, $\epsilon/\wc =10^{-4}$, and
$\w_\mrm{max} = 10 \wc$. }

\label{fig:fkBias}
\end{figure}

\begin{figure}[tb]
\includegraphics[width=0.99\columnwidth]{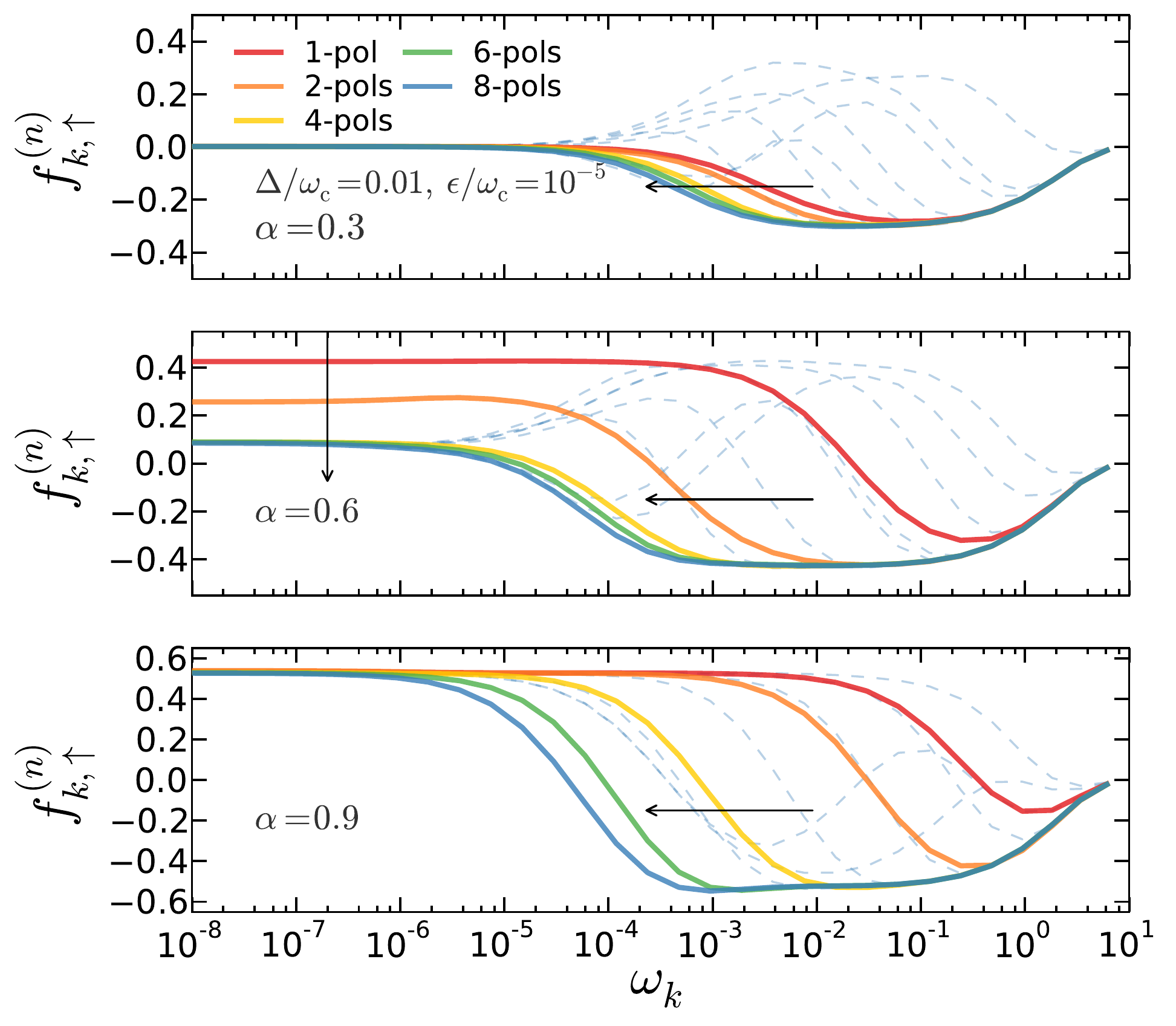}
\caption{Polaronic displacement $f_{k,\uparrow}^{(1)}$ computed for an
  increasing number of coherent states, $N_\mrm{pols}=1,2,4,6,8$ (full curves,
  from right to left) in Eq.~(\ref{eq:GSbias}). Parameters: $\Delta/\w_c=0.01$
  and $\epsilon/\w_c=10^{-5}$, $\alpha=0.3$ (top), $\alpha=0.6$ (middle), and
  $\alpha=0.9$ (bottom).  The multi-node displacements for $N_\mrm{pols}=8$ are
shown as dotted curves. As in Fig.~\ref{fig:fkBias}, an exponential high-energy
cutoff was used in the spectral density.}
\label{fig:Fk1BiasNpols}
\end{figure}

In Fig.~\ref{fig:fkBias} we show the variationally-determined oscillator displacements 
calculated using the biased multi-polaron ansatz of Eq.~\eqref{eq:GSbias} for $N_\mrm{pols} = 6$ 
coherent states.
Again, at high energy, $\w_k \gg \Delta_{R}$, all displacements smoothly merge 
with the main polaron $f_{k,\ua/\da}^{(1)}$~(red solid lines). 
At intermediate frequencies, extra nodal structure emerges, due to the energetic
gains associated to quantum tunneling. Finally, all displacements converge to the
same finite value at vanishing energy, an effect of the finite applied bias.
These fully optimized variational parameters follow nearly quantitatively the
form of the parametrized displacements in Eq.~\eqref{eq:fkParamBias}, in
agreement with energetic considerations.
As in the unbiased case, the main polaronic displacement differs
quantitatively from the single-polaron (SH) prediction, with the renormalized
tunneling scale $\Delta_R$ again pushed downwards as antipolarons are added into
the trial state. This is illustrated for $f_{k,\uparrow}^{(1)}$ in
Fig.~\ref{fig:Fk1BiasNpols} at various dissipation strengths. Additionally,
notice the change in behavior at vanishing energy, seen most clearly for
$\alpha=0.6$, where the finite displacement due to the applied bias is
suppressed as antipolarons are added to the wavefunction. The latter
occurs because for these parameters the spin is partially polarized and poised
to abruptly switch from being unpolarized to fully polarized (see next section),
thus making the state especially sensitive to the number of polarons included in
the wavefunction ansatz.

\subsubsection{Magnetization.} 
\begin{figure}[h]
	\centering
\includegraphics[width=0.95\columnwidth]{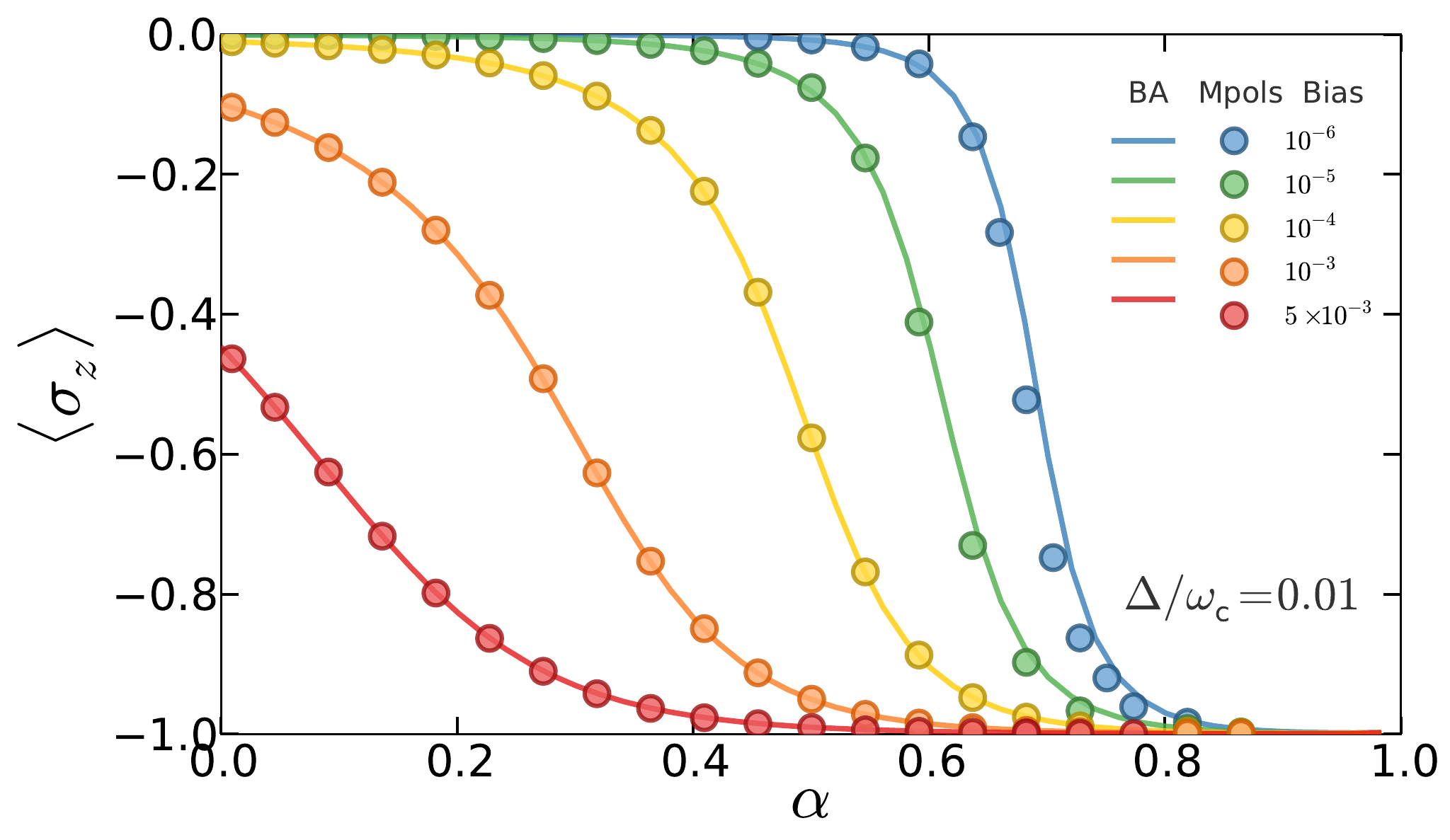}
\caption{Two-level population difference (magnetization) $\la\spin{z}\ra$ as a
  function of dissipation strength $\alpha$ for increasing values of the bias
  field $\epsilon/\omega_c$ (top to bottom) with $\Delta/\wc=0.01$: circles
  (labeled Mpols) mark results from the multi-polaron ansatz,
  Eq.~\eqref{eq:GSbias}, and solid lines are Bethe Ansatz~(BA) results.  All
  circles were calculated using only $N_{\rm pols} = 6$ polarons, except for the
  case $\epsilon/\wc =10^{-6}$ where for better (but not yet complete)
  convergence $N_{\rm pols} = 8$ polarons were used. 
}
\label{fig:SzBias}
\end{figure}

In Fig.~\ref{fig:SzBias} we investigate the behavior of the two-level
population difference (or qubit magnetization) as a function of spin-bath
coupling strength $\alpha$ for several different bias values. 
As a benchmark, we use exact Bethe Ansatz results, which are valid in the scaling 
limit~($\Delta/\wc\ll 1$)~\cite{LeHurReview} and for a (soft) exponential bosonic
cutoff in the spectral density, $J(\w) = 2\pi \alpha \w e^{-\w/\wc}\theta(\w_{\mrm{max}}-\w)$, which we also employ in our
biased multi-polaron calculations in this section (with
$\w_{\mrm{max}}=10 \w_c$). Indeed, because the Bethe Ansatz solution relies 
on bosonization identities to map the spin-boson model to an exactly solvable fermionic model, we need to
be careful to respect the natural cutoff associated with bosonization in order to
make quantitative comparisons.
This results in excellent agreement between the Bethe Ansatz formula~(solid curves) 
and our multi-polaron results~(circles). Here, we used $N_{\rm pols} = 6$ polarons, except
for the smallest bias $\epsilon/\wc = 10^{-6}$, where $N_{\rm pols} = 8$ polarons were required to
achieve better convergence. Typically, the number of coherent states required
for convergence in the expansion~(\ref{eq:GSbias}) increases at stronger dissipation.
We emphasize again that SH theory for the biased spin-boson model fails 
to correctly predict the smooth crossover in magnetization as a function of
dissipation,~\cite{Nazir} giving rise instead to an unphysical ``jump" for certain bias 
values, while in our ansatz no such discontinuous behavior appears.

\begin{figure}[t]
	\centering
\includegraphics[width=0.95\columnwidth]{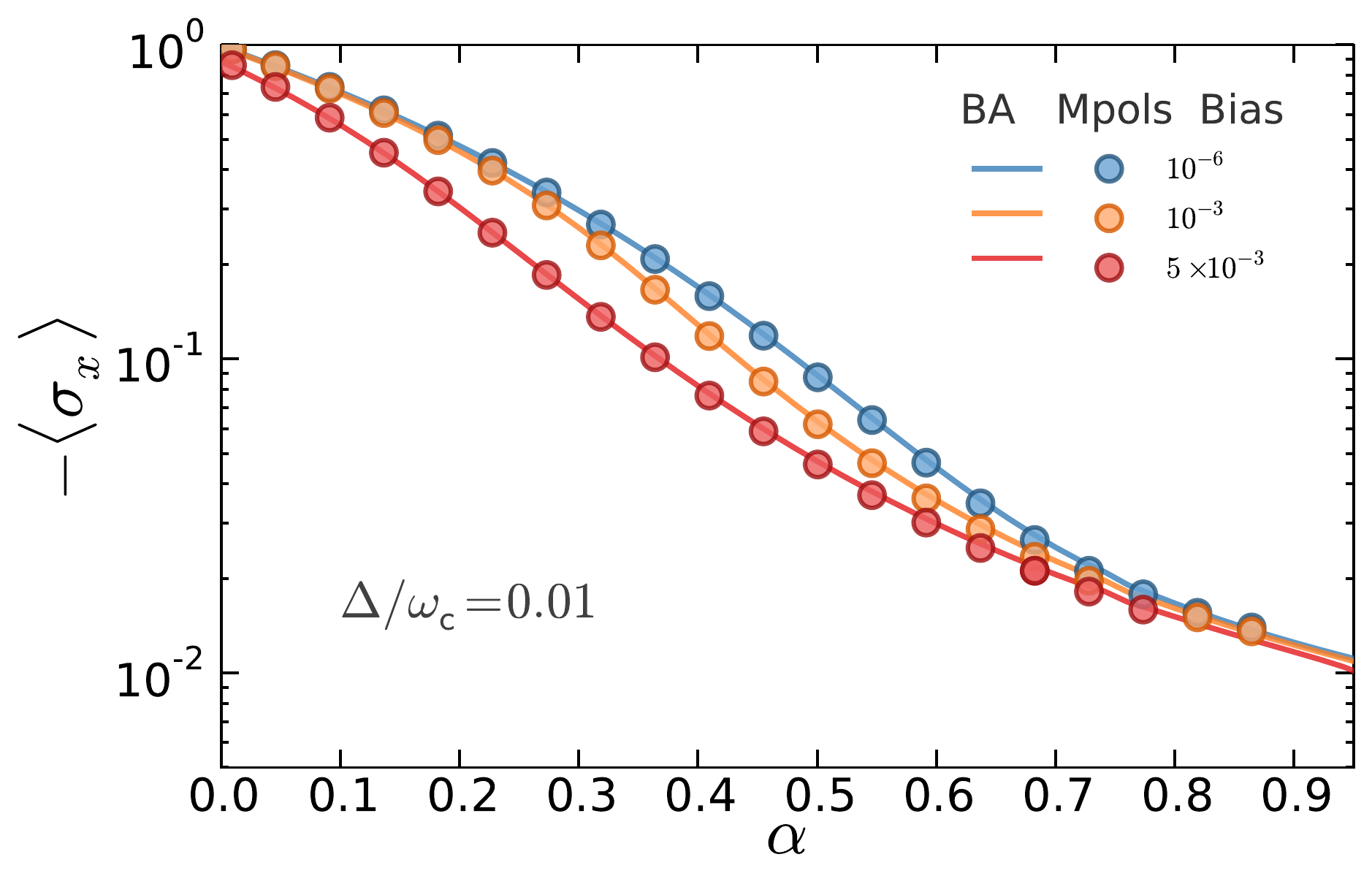}
\caption{Ground state spin coherence $\la\spin{x}\ra$ as a function of
dissipation $\alpha$ for increasing values of the bias field $\epsilon/\omega_c$ (top to
bottom), as calculated using the multi-polaron~(Mpols) ansatz (circles) and 
the exact Bethe Ansatz~(solid lines). Parameters are the same as in Fig.~\ref{fig:SzBias}.
}
\label{fig:SxBias}
\end{figure}

\subsubsection{Spin coherence.}
The spin coherence $\la\spin{x}\ra$ shown in Fig.~\ref{fig:SxBias}
is also in excellent agreement with the Bethe Ansatz results, verifying again 
that we obtain the correct underlying description of the ground state wavefunction from our multi-polaron
ansatz.
While the magnetization $\la\spin{z}\ra$ depends only on fixed point properties
(it is a scaling function of $\epsilon/\Delta_R)$, the Bethe
Ansatz expression for the spin coherence~\cite{LeHurReview} contains both fixed point contributions and a 
non-universal correction of the order of $\Delta/\wc$.
Looking at the essentially perfect agreement in Fig.~\ref{fig:SxBias}, it is clear that our
multi-polaron ansatz is flexible enough to successfully capture {\it both universal and
non-universal features} of the ground state properties of the biased spin-boson
model. Our results are also consistent with the Bethe Ansatz
prediction that the spin coherence should be $\la\spin{x}\ra\sim \Delta/\wc$ at
the quantum critical point, for $\epsilon\to0$.~\cite{Ponomarenko,LeHurReview}

\subsubsection{Convergence of the multi-polaron expansion.}
\begin{figure}[t]
	\centering
\includegraphics[width=1.0\columnwidth]{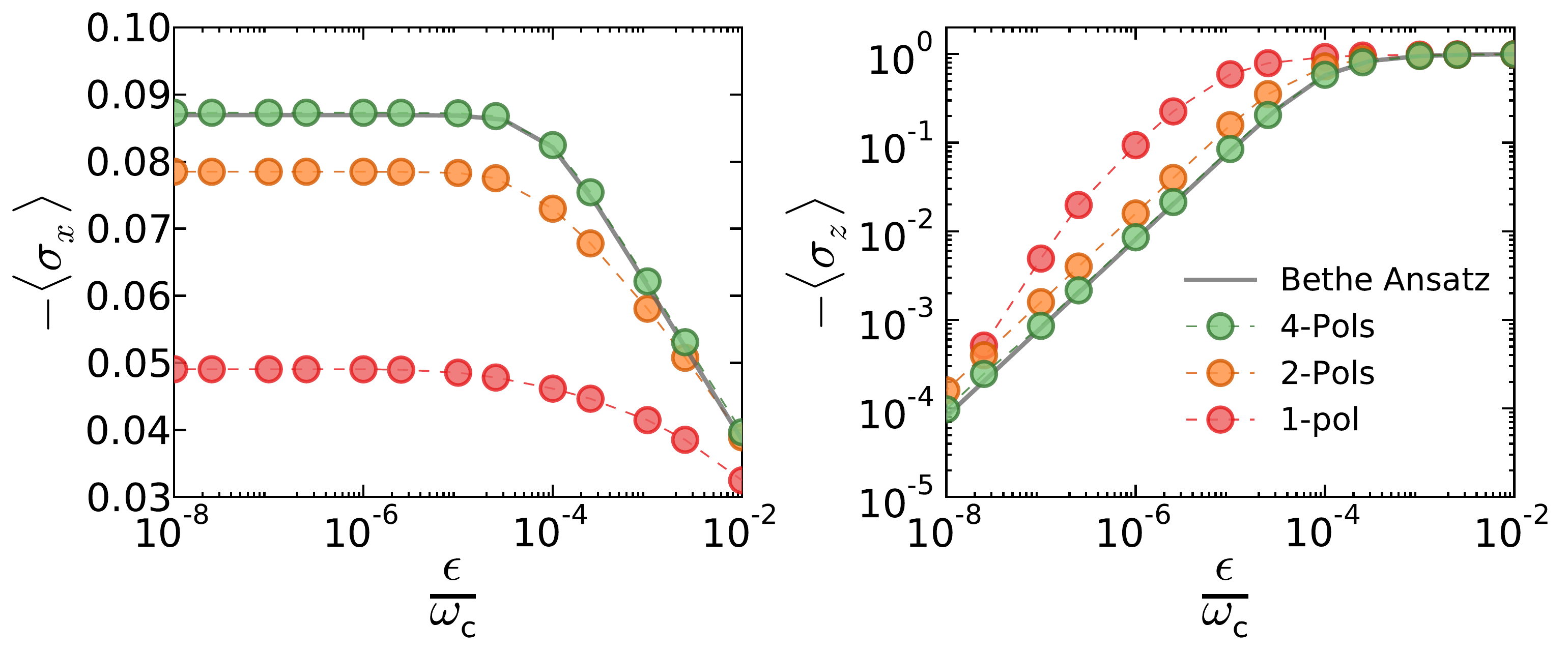}
\includegraphics[width=1.0\columnwidth]{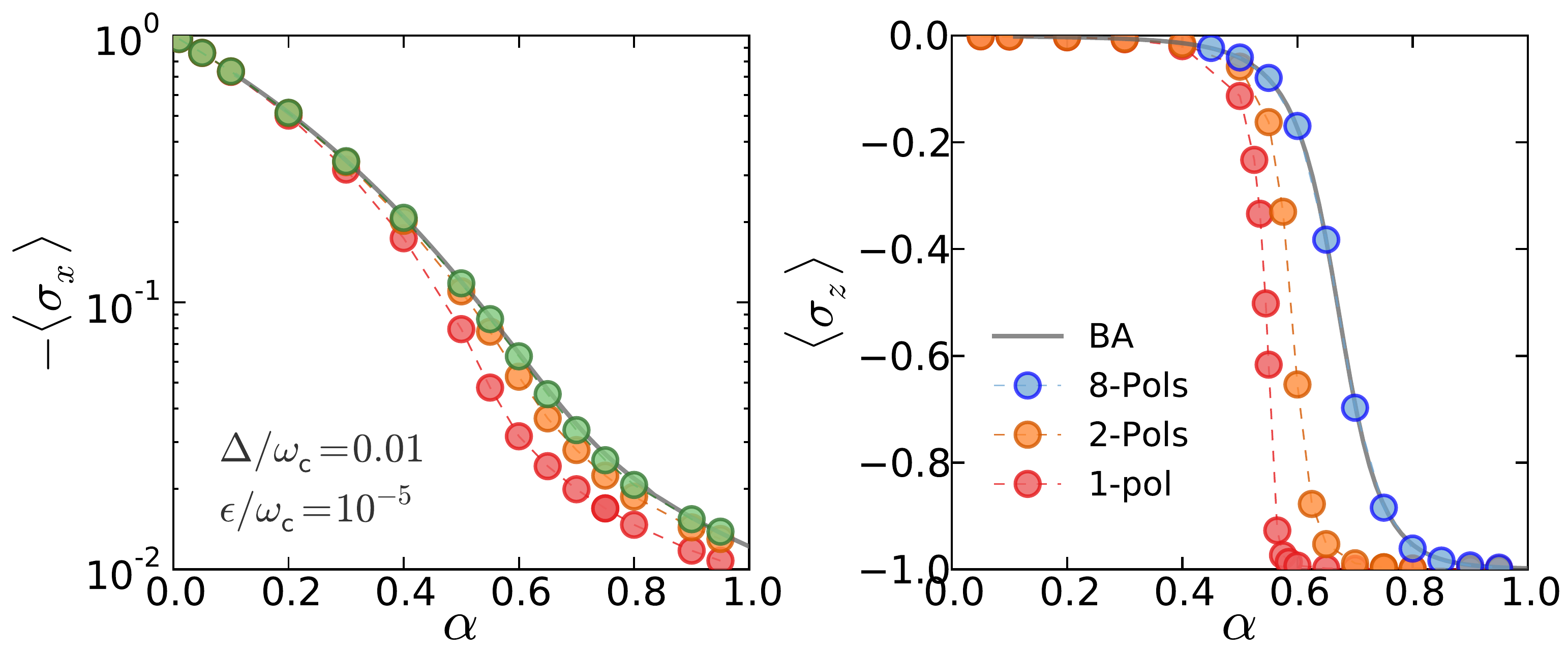}
\caption{Ground state spin coherence $\la\spin{x}\ra$ and magnetization
	$\la\spin{z}\ra$ as a function of bias $\epsilon$ (in units of $\omega_c$) for $\alpha=0.5$ (top
panels), and as a function of $\alpha$ for $\epsilon/\wc=10^{-5}$ (bottom panels).
The multi-polaron curves (joined circles) are calculated with $N_\mrm{pols}=1,2,4,8$, and
compared to the Bethe Ansatz (solid lines).
}
\label{fig:SxSzTpt}
\end{figure}

Finally, we shall investigate 
the convergence properties
of our coherent state 
expansion~(\ref{eq:GSbias}) in the biased case.
In Fig.~\ref{fig:SxSzTpt} we show the bias dependence of the spin 
coherence and magnetization (top panels) for $\alpha=0.5$ (the Toulouse point),
using an increasing number of polarons.
These plots demonstrate rapid convergence with only $N_\mrm{pols}=4$ coherent states;
in contrast, large quantitative deviations exist for the asymmetric single-polaron SH
ansatz, 
Eq.~(\ref{eq:aSHbias}) (labeled~$1$-pol).
For small bias, $\epsilon \ll \Delta_{R}$, the spin coherence $\la \spin{x}\ra$
saturates, while the magnetization $\la \spin{z}\ra$ behaves linearly as
$(2/\pi) (\epsilon/\Delta_{R})$ in accordance with the Bethe Ansatz
calculations. For large bias $\epsilon \gg \Delta_{R}$, the spin coherence decreases due to the complete
saturation $\la\spin{z}\ra\to -1$.
The two lower panels in Fig.~\ref{fig:SxSzTpt} show the spin coherence and
magnetization for fixed $\epsilon/\wc=10^{-5}$, as a function of dissipation strength $\alpha$.
We note here that the coherence obtained with the SH state
is more accurate at finite $\epsilon$ than in the $\epsilon=0$ limit (see Fig.~\ref{fig:Sx}), because
tunneling effects are suppressed in the presence of a finite bias. 
However, the magnetization shows a very abrupt jump as a function of increasing $\alpha$, 
that is correctly smoothed out in the converged multi-polaron solution.

\section{Conclusions}

We have 
further developed a coherent state expansion for the dissipative
quantum two-level system (the spin-boson model), and proposed a simple extension of the method to
include the effect of arbitrary magnetic fields (transverse and longitudinal to the bath).
Excellent agreement is found with controlled numerical renormalization group calculations
and the exact Bethe Ansatz solution, at relatively low computational
expense. We have also constructed a simple and appealing physical picture in
terms of displacements of the many-body wavefunction of the bath, which turns out 
to be strongly entangled at large dissipation. Several interesting directions of research 
are opened with the development
of this new methodology. First, one can investigate the nature of a quantum
critical wavefunction by considering the sub-Ohmic spin-boson model, which shows
a second-order critical point (the Ohmic case falls into the Kosterlitz-Thouless
class). More ambitious is the extension to increasingly complex sub-systems (e.g. higher
spins or several states within the sub-system) and several baths, a problem relevant, for example, to 
exciton dynamics in biological systems.~\cite{Plenio} Finally, the issue of dynamical
quenches~\cite{Leggett} or photon
scattering in superconducting circuit setups~\cite{LeHurJosephson,GoldsteinJosephson,Wilhelm} 
may be considered by a time-dependent version
of our variational coherent state wavefunction.

\acknowledgments
We wish to thank N. Roch and D. P. S. McCutcheon for stimulating discussions,
and Z. Blunden-Codd for detailed comments on the manuscript.
We thank the Grenoble Nanoscience Foundation for
funding under RTRA contract CORTRANO. 
The work at Duke was supported by the US DOE, Division of 
Materials Sciences and Engineering, under Grant No.\,DE-SC0005237.
A.N. thanks The University of Manchester and Imperial College London for support. A.W.C. acknowledges support from 
the Winton Program for the Physics of Sustainability.

\appendix

\section{Details on the bosonic NRG calculations}
\label{AppendixA}
\subsubsection{NRG procedure.} In order to check our unbiased multi-polaron ansatz,
Eq.~\eqref{eq:GS}, we perform accurate numerical renormalization group~(NRG)
calculations for the Ohmic spin-boson model. In the biased case, due to 
trickier convergence issues with the NRG, we compared our results instead to the Bethe Ansatz.
We followed the standard and simplest implementation of the NRG method, which was 
initially introduced for fermionic Kondo problems.~\cite{NRG-RMP08}
The key idea of NRG is scale separation that results from the logarithmic
discretization of the bosonic energy band. The model is subsequently solved
by iterative diagonalization of only a small number of degrees of freedom at each
NRG step. 

We start with the continuum expression for the spin-boson model:
\be
\mc{H} = \f{\Delta}{2} \spin{x} + \int_{0}^{\wc} \mrm{d}\veps  \, \veps\,  a_{\veps}^{\dg}
a_{\veps} -\f{\spin{z}}{2} \int_{0}^{\wc} \mrm{d}\veps \,	h(\veps) \Le(
a_\veps + a_\veps^\dg\Ri),
\label{eq:nrgH}
\ee
where $h(\veps)$ is related to the spectral density,~Eq.~\eqref{eq:sden}, by $J(\veps)= \pi
h(\veps)^2$. Next, the bosonic bath is discretized logarithmically in the
interval $\Le[0, \wc\Ri]$, with $d_n=\w_c\Lambda^{-n} (1-\Lambda^{-1})$  the width
of each interval and $\Lambda$ the Wilson parameter, typically $\Lambda=2$. 
In the next step, the bosonic fields are expanded in Fourier modes in each interval: 
\be
a_\veps^\dg = \sum_{n, p} \f{1}{\sqrt{d_n}}  e^{i \f{2 \pi p \veps}{d_n}}
a_{n, p}^\dg,
\ee
with $p$ an integer.
The Hamiltonian~[Eq.~\eqref{eq:nrgH}] is then re-expressed in the Fourier basis,
and because the $p \ne 0$ terms are decoupled from the impurity,
they are dropped, which is vindicated for large enough $\Lambda$.
We then end up with the so-called star Hamiltonian in terms of the $p = 0$ mode only:
\be
\mc{H}_\mrm{star} = \f{\Delta}{2} \spin{x} 	+ \sum_{n=0}^{\infty} \xi_n \
a_n^\dg a_n - \f{\spin{z}}{2 \sqrt{\pi}} \sum_{n=0}^{\infty} \gamma_n \ (a_n +
a_n^\dg), 
\label{eq:nrgStarH}
\ee
where the energy of each bosonic mode and the impurity coupling strength are given, respectively, by
\begin{eqnarray}
\xi_n &= &\f{1}{\gamma_n^2} \int_{\Lambda^{-(n+1)}
\wc}^{\Lambda^{-n} \wc}
J(\w)\, \w\,  \mrm{d} \w, \\
\gamma_n^2 &=& \int_{\Lambda^{-(n+1)} \wc}^{\Lambda^{-n} \wc} J(\w)
\,  \mrm{d} \w. 
\end{eqnarray}
For an Ohmic spectral density (with hard cut-off), these parameters read
\be
\xi_n = \f{2}{3} \f{1-\Lambda^{-3}}{1-\Lambda^{-2}} \wc \Lambda^{-n}, \quad
\gamma_n^2 = \pi \alpha \wc^2 \Le(1-\Lambda^{-2}\Ri) \Lambda^{-2 n}. 
\ee
The next, but not obligatory, step of NRG is to represent the star Hamiltonian
by a semi-infinite chain:
\begin{align}
	\mc{H}_\mrm{chain} =& \f{\Delta}{2}\spin{x} -
\f{\spin{z}}{2}\sqrt{\f{\eta_0}{\pi}} \Le(b_0 + b_0^\dg \Ri) + \nn \\
&+ \sum_{n=0}^{\infty} \Le[ \epsilon_n b_n^\dg b_n + t_n (b_n^\dg b_{n+1} + {\rm h.
c.})\Ri],
\label{eq:nrgChainH}
\end{align}
where $\eta_0=\int J(\w)\, \mrm{d}\w$, and $\epsilon_n$ and $t_n$ are
the on-site energy and the hopping amplitude between different sites of the
chain, respectively, which can readily be calculated.
Within the chain representation, the spin is only coupled to the
first site, and all other bosonic sites are coupled successively
to each other with exponentially decreasing hopping energies. The bosonic NRG
procedure starts by diagonalizing the coupled system of spin and the first
chain site, and the renormalization follows by adding extra sites successively,
while truncating the Hilbert space at each step (justified by
the property of scale separation). This ensures stability of the NRG, which
can then reach very small energy scales in a linear effort. Typically, in all our
NRG calculations we construct the local Hilbert space of a given site with up
to $N_\mrm{b} = 10$ bosons, and we truncate the total Hilbert space at each
stage to 200 states (the matrices to be diagonalized then do not exceed 
$2000\times2000$ in size).

\subsubsection{Convergence speed-up to the $\Lambda\to1$ limit.}
In principle, the NRG is exact for a dense spectrum $\Lambda\to1$,
but in that case the scale separation breaks down and the truncation
becomes unmanageable. 
On the other hand, having a larger $\Lambda$ improves scale separation, 
but leads to a poor approximation of the bath states.
A compromise must therefore be found, and typically calculations with
$\Lambda=2$ ensure good quality NRG results. An easy improvement
to the convergence can be made by adjusting the dissipation strength in
an appropriate manner. One possibility is to fine tune $\alpha$ in order
to identify the universal scale of the continuum limit with the slightly
different value obtained on the Wilson discretization. Constraining
$\Delta_R^{\mrm{\Lambda=1}} = \Delta_R^{\mrm{NRG}}$ gives: 
\begin{eqnarray}
\nonumber
\Delta \exp\left[ -2\int \frac{J(\w)}{(\w+\Delta_R)^2}\right] &=& 
\Delta \exp\left[ -2\sum_{n=0}^{+\infty} \frac{(\gamma_n)^2}{(\xi_n+\Delta_R)^2}
\right] \\
\nonumber
\Rightarrow
\int_0^{\w_c} \frac{\alpha}{(\w+\Delta_R)^2} &=&
\frac{9}{8\ln \Lambda}
\frac{(1-\Lambda^{-2})^3}{(1-\Lambda^{-3})^2} \times\\
\nonumber
&& \times
\sum_{n=0}^{+\infty}
\frac{\alpha^{\mrm{NRG}}}{1+\Delta_R \Lambda^n}.
\end{eqnarray}
In the scaling limit $\Delta\ll\w_c$, $\Delta_R$ is very small, and both
the sum and integral behave logarithmically. We can then relate the
dissipation strengths as
\begin{equation}
\alpha = \frac{9}{8\ln \Lambda}
\frac{(1-\Lambda^{-2})^3}{(1-\Lambda^{-3})^2} \alpha^{\mrm{NRG}} \simeq 0.894
\alpha^{\mrm{NRG}},
\end{equation}
for $\Lambda=2$.
Thus, the NRG dissipation strength $\alpha^{\mrm{NRG}}$ must be renormalized 
by about $10\%$ if one wants to make quantitative comparisons to NRG
calculations performed with a Wilson parameter $\Lambda=2$.
We emphasize that this prescription is not rigorous, as it depends slightly
on which physical quantity is used in making the identification of the coupling
$\alpha$.

\section{Computing the entanglement entropy}
\label{AppendixB}

\subsection{From the coherent state expansion}

We consider first the spin entanglement entropy.
From the total density matrix associated with the pure ground state,
$\rho_{\mathrm{tot}}=\big|\Psi\big>\big<\Psi\big|$,
we deduce the reduced density matrix of the spin alone, 
obtained by tracing over all environmental modes:

$\rho_{\mathrm{spin}} = \mathrm{Tr}_{\mathrm{env}}
\big|\Psi\big>\big<\Psi\big|$.
\begin{widetext}
In the spin state basis ($\sigma=\uparrow,
\downarrow$) this reads:
\begin{equation}
\rho_{\mathrm{spin}}=
\frac{1}{\big<\Psi\big|\Psi\big>}
\left[
\begin{array}{cc}
+\sum_{n,m} C_n C_m
\big<+f^{(n)}\big|+f^{(m)}\big>
&- \sum_{n,m} C_n C_m
\big<+f^{(n)}\big|-f^{(m)}\big> \\
- \sum_{n,m} C_n C_m
\big<-f^{(n)}\big|+f^{(m)}\big> 
&+ \sum_{n,m} C_n C_m
\big<-f^{(n)}\big|-f^{(m)}\big>
\end{array}
\right].
\end{equation}
Obviously $\mathrm{Tr}_{\mathrm{spin}} \rho_{\mathrm{spin}}=1$,
but $\mathrm{Tr}_{\mathrm{spin}} \rho_{\mathrm{spin}}^2<1$ as
we now have a mixed state. This allows us to define the $\rho^2$-entropy
(qualitatively similar to the usual von Neumann entropy), 
$ S_{\mathrm{spin}}= 1-\mathrm{Tr} [\rho_{\mathrm{spin}}^2]$.
Using symmetry properties of coherent states, we readily find a
compact expression in terms of the trial wavefunction:
\begin{equation}
\label{Sspinfinal}
S_{\mathrm{spin}}= 1-\frac{2}{\left[\big<\Psi\big|\Psi\big>\right]^2}
\left[
\left( \sum_{n,m} C_n C_m
e^{-\frac{1}{2}\sum_{q} (f^{(n)}_q-f^{(m)}_q)^2}
\right)^2
+
\left( \sum_{n,m} C_n C_m
e^{-\frac{1}{2}\sum_{q} (f^{(n)}_q+f^{(m)}_q)^2}
\right)^2
\right].
\end{equation}

We now consider the reduced density matrix, 
 Eq.~(\ref{rhospink}), 
obtained by tracing out all 
modes except the qubit degree of freedom together with an {\it arbitrary} bath
mode with given quantum number $k$.
Its matrix elements in the combined qubit ($\sigma=\uparrow,\downarrow$) and
Fock basis of mode-$k$ ($\big|l\big>_k$, with $l=0,\ldots,\infty$) are:
\begin{eqnarray} 
[\rho_{\mathrm{spin}+k}]_{\sigma,\sigma';l,l'} &=& \big<\Psi\big| 
\Big[\big|l\big>_k \big|\sigma\big> 
\big<\sigma'\big| \!{\phantom{\big>}}_k\big<l'\big|\Big]
\big|\Psi\big>  \\
\nonumber
&=& \frac{1}{\big<\Psi\big|\Psi\big>}
\sum_{n} C_n 
\Big[ \big<+f^{(n)}\big|\big<\uparrow\big|-
\big<-f^{(n)}\big|\big<\downarrow\big|\Big]
\big|l\big>_k \big|\sigma\big>
\sum_{m} C_m 
\!{\phantom{\big>}}_k \big<l'\big| \big<\sigma\big|
\Big[ \big|+f^{(m)}\big>\big|\uparrow\big>-
\big|-f^{(m)}\big>\big|\downarrow\big>\Big]\\
\nonumber
&=& \frac{1}{\big<\Psi\big|\Psi\big>}
\sum_{n,m} C_n C_m \left(
\delta_{\sigma\uparrow}\delta_{\sigma'\uparrow}
+\delta_{\sigma\downarrow}\delta_{\sigma'\downarrow}
\right)
\!{\phantom{\big>}}_{q\neq k} \big<f^{(n)}\big|f^{(m)}\big>_{q\neq k} 
\!{\phantom{\big>}}_{k} \big<f^{(n)}\big|l\big>_{k} 
\!{\phantom{\big>}}_{k} \big<l'\big|f^{(m)}\big>_{k} \\
\nonumber
&&- \frac{1}{\big<\Psi\big|\Psi\big>}
\sum_{n,m} C_n C_m \left(
\delta_{\sigma\uparrow}\delta_{\sigma'\downarrow}
+\delta_{\sigma\downarrow}\delta_{\sigma'\uparrow}
\right)
\!{\phantom{\big>}}_{q\neq k} \big<f^{(n)}\big|-f^{(m)}\big>_{q\neq k} 
\!{\phantom{\big>}}_{k} \big<f^{(n)}\big|l\big>_{k} 
\!{\phantom{\big>}}_{k} \big<l'\big|-f^{(m)}\big>_{k} \\
\nonumber
&=& \frac{1}{\big<\Psi\big|\Psi\big>}
\sum_{n,m} C_n C_m \left(
\delta_{\sigma\uparrow}\delta_{\sigma'\uparrow}
+\delta_{\sigma\downarrow}\delta_{\sigma'\downarrow}
\right)
e^{-\frac{1}{2}\sum_{q\neq k} (f^{(n)}_q-f^{(m)}_q)^2}
\frac{[f^{(n)}_k]^l[f^{(m)}_k]^{l'}}{\sqrt{l! l'!}}
e^{-\frac{1}{2} (f^{(n)}_k)^2 -\frac{1}{2} (f^{(m)}_k)^2}\\
\nonumber
&& - \frac{1}{\big<\Psi\big|\Psi\big>}
\sum_{n,m} C_n C_m \left(
\delta_{\sigma\uparrow}\delta_{\sigma'\downarrow}
+\delta_{\sigma\downarrow}\delta_{\sigma'\uparrow}
\right)
e^{-\frac{1}{2}\sum_{q\neq k} (f^{(n)}_q+f^{(m)}_q)^2}
\frac{[f^{(n)}_k]^l[-f^{(m)}_k]^{l'}}{\sqrt{l! l'!}}
e^{-\frac{1}{2} (f^{(n)}_k)^2 -\frac{1}{2} (f^{(m)}_k)^2},
\end{eqnarray} 
where the last equation results from straightforward coherent state algebra.
We finally obtain the spin+mode entropy:
\begin{eqnarray}
\label{Sspink}
S_{\mathrm{spin}+k} & = & 1-\mathrm{Tr} [\rho_{\mathrm{spin}+k}^2]
= 1-\sum_{\sigma,\sigma';l,l'} (\rho_{\sigma,\sigma';l,l'})^2 \\
\label{Sspinmodefinal}
&=& 1 - \frac{2}{(\big<\Psi\big|\Psi\big>)^2} \sum_{l,l'}
\Bigg[ \left(\sum_{n,m} C_n C_m 
e^{-\frac{1}{2}\sum_{q\neq k} (f^{(n)}_q-f^{(m)}_q)^2}
\frac{[f^{(n)}_k]^l[f^{(m)}_k]^{l'}}{\sqrt{l! l'!}}
e^{-\frac{1}{2} (f^{(n)}_k)^2 -\frac{1}{2} (f^{(m)}_k)^2}\right)^2\\
\nonumber
&& + \left(\sum_{n,m} C_n C_m 
e^{-\frac{1}{2}\sum_{q\neq k} (f^{(n)}_q+f^{(m)}_q)^2}
\frac{[f^{(n)}_k]^l[-f^{(m)}_k]^{l'}}{\sqrt{l! l'!}}
e^{-\frac{1}{2} (f^{(n)}_k)^2 -\frac{1}{2} (f^{(m)}_k)^2}\right)^2
\Bigg].
\end{eqnarray}
Eqs.~(\ref{Sspinfinal}) and (\ref{Sspinmodefinal}) were used
to compute the entanglement mode spectroscopy in the bottom panel of Fig.~\ref{fig:Rh2}.


\subsection{From the NRG}

The strategy used for the NRG entanglement
entropy computation (shown in Sec.~\ref{sec:entropy}) relies on first obtaining the reduced density matrix
$\rho_{\mathrm{spin}+k}$, which acts within the 
subspace spanned by the qubit and a single bosonic mode $k$. We start by 
defining the joint spin and Fock projection operator
$O_{\sigma_i;m,m'}^{(k)}= \sigma_i |m\big>_{k\;k}\big<m'|$, so that
matrix elements of the ground state density matrix simply read
\begin{equation}
\rho_{\sigma_i,m,m'}^{(k)} = \big<\Psi| O_{\sigma_i;m,m'}^{(k)} |\Psi\big> \;.
\end{equation}
This quantity is a ground state average, hence readily computable by
letting the operator $O_{\sigma_i;m,m'}^{(k)}$ evolve along the complete
NRG flow. Tracing the squared matrix $\rho_{\sigma_i,m,m'}^{(k)}$
allows us to obtain the desired entanglement entropy, shown in the top panel of Fig.~\ref{fig:Rh2}.

\section{Computing the energy variance}
\label{AppendixC}

Here we provide an explicit formula for the energy variance in terms of the
coherent state expansion~(\ref{eq:GS}). We start by squaring the spin-boson
Hamiltonian~(\ref{eq:SB}), and use normal ordering of the bosonic operators:
\begin{eqnarray}
\nonumber
\mc{H}^2 &=& \Delta \sigma_x \sum_k \w_k \bkd \bk  + \frac{\Delta^2}{4} + \sum_k
\w_k^2 \bkd \bk + \sum_{k,k'} \w_k \w_{k'} \bkd b^{\dagger}_{k'} \bk b^{\phantom{\dagger}}_{k'}
+ \frac{1}{4}\sum_k g_k^2 + \frac{1}{4}\sum_{k,k'} g_k g_{k'} 
(\bkd b^{\dagger}_{k'} 
+ \bk b^{\phantom{\dagger}}_{k'} + \bkd
b^{\phantom{\dagger}}_{k'} + b^{\dagger}_{k'} \bk)\\
&& - \frac{\sigma_z}{2} \sum_k g_k \w_k (\bk+\bkd) 
- \sigma_z \sum_{k,k'} g_{k'} \w_k (\bkd b^{\phantom{\dagger}}_{k'} \bk
+ \bkd b^{\dagger}_{k'} \bk).
\end{eqnarray}
Coherent state algebra then enables us to compute the expectation value for the
trial state~(\ref{eq:GS}):
\begin{eqnarray}
\nonumber
\big<\mc{H}^2\big> &=&  
- 2 \sum_{n,m} C_n C_m \big<f^{(n)}|-f^{(m)}\big>
\Delta \sum_k \w_k f_k^{(n)} f_{k}^{(m)}+
2 \sum_{n,m} C_n C_m
\big<f^{(n)}|f^{(m)}\big> \Bigg\{
\frac{\Delta^2}{4} 
+ \sum_k \w_k^2 f_k^{(n)} f_k^{(m)}\\
\nonumber
&&
+ \sum_{k,k'} \w_k \w_{k'} f_k^{(n)} f_{k'}^{(m)} f_{k'}^{(n)} f_k^{(m)}
+ \frac{1}{4}\sum_k g_k^2 
+ \frac{1}{4}\sum_{k,k'} g_k g_{k'} 
\left(f_k^{(n)}+f_k^{(m)}\right)\left(f_{k'}^{(n)}+f_{k'}^{(m)}\right)\\
&&
- \frac{1}{2} \sum_k g_k \w_k 
\left(f_k^{(n)}+f_k^{(m)}\right) 
- \sum_{k,k'} g_{k'} \w_k
f_k^{(n)} f_k^{(m)} \left( f_{k'}^{(n)}+ f_{k'}^{(m)} \right) 
\Bigg\}.
\label{Hsquare}
\end{eqnarray}
The energy variance $\big<\mc{H}^2\big>-\big<\mc{H}\big>^2$ can finally be 
obtained by combining Eq.~(\ref{Hsquare}) above and Eq.~(\ref{eq:Menergy}) 
for the ground state energy.
\end{widetext}

\end{document}